\documentclass[twocolumn,showpacs,pra]{revtex4}
\usepackage{epsfig}
\usepackage{dcolumn}
\usepackage{bm}

\begin{document}

\title{Excitation spectra of a $^3$He impurity on $^4$He clusters}
\author{S. Fantoni}

\author{R. Guardiola}
\altaffiliation{On leave of absence from
Departamento de F\'{\i}sica At\'omica y Nuclear,
Facultad de F\'{\i}sica, E-46100-Burjassot, Spain}
\affiliation{SISSA and INFM DEMOCRITOS National Simulation Center, 
Via Beirut 2-4, I-34014 Trieste, Italy}

\author{J. Navarro}
\affiliation{
IFIC (CSIC-Universidad de Valencia),
Apartado Postal 22085, E-46071-Valencia, Spain}

\date{\today}

\begin{abstract}
The diffusion Monte Carlo technique is used to calculate and analyze the
excitation spectrum of a single $^3$He atom bound to a cluster with $N$
$^4$He atoms, with the aim of establishing the most adequate filling
ordering of single-fermion orbits  to the  mixed clusters with a large number 
of $^3$He atoms. The resulting ordering looks like the rotational spectrum 
of a diatomic molecule, being classified only by the angular momentum of 
the level, although vibrational-like excitations  appear at higher energies 
for sufficiently large $N$.
\end{abstract}


\maketitle

\section{Introduction}
The study of isotopic $^4$He-$^3$He mixed clusters is attracting a growing
interest in recent years. From the experimental viewpoint, the diffraction
of clusters from a transmission grating~\cite{Scho94} has opened new
perspectives in the detection and identification of small clusters. There is
no evidence for the existence of the dimer $^4$He$^3$He, but clusters
$^4$He$_N$$^3$He with $N=2,3,4$ have been definitely detected~\cite{Scho96}.
It seems possible at present to resolve clearly the clusters of mass up to
about 25 amu~\cite{Korn03}. As the weakly van der Waals He-He interaction is 
isotope independent, the properties of such mixed clusters are determined 
solely by quantal effects, namely the different zero-point motion and the 
different statistics of the two isotopes. It turns out that helium clusters 
are weakly bound systems, and the lighter ones are challenging for microscopic 
theoretical methods.

The stability of small mixed clusters has been the object of several recent
microscopic studies. Guardiola and Navarro have investigated clusters 
containing up to eight $^4$He atoms and up to 20 $^3$He atoms, based on both
a Variational Monte Carlo (VMC) wave function~\cite{Guar02,Guar03} and the 
diffusion Monte Carlo (DMC) method~\cite{Guar03a} in the fixed-node
approximation. Bressanini and collaborators~\cite{Bres00,Bres02,Bres03} have
considered clusters with up to 17 $^4$He and up to three $^3$He atoms, by
means of the DMC method, also in the fixed node approximation.

The DMC description is based in an importance sampling wave function which
plays a triple role: it controls the variance of the ground-state energy, it
carries on the quantum numbers and other properties of the considered cluster,
and it specifies the nodal (or set of nodal) surfaces. In particular, in
Refs.~\cite{Guar02,Guar03,Guar03a} the antisymmetry required for fermions has 
been taken into account by means of two Slater determinants, one for each spin 
orientation, which have been built up in terms of harmonic polynomials in the 
Cartesian coordinates of each fermion. Moreover, a harmonic-oscillator (HO) 
ordering of the fermionic shells has been assumed. Although this seems a 
reasonable hypothesis, supported by the findings of density-functional 
calculations in medium sized mixed droplets~\cite{Barr97}, from a microscopic 
point of view there are no conclusive {\it a priori} arguments in favor of 
such an ordering. For instance, to describe mixed systems with three $^3$He 
atoms, a configuration with total angular momentum $L=0$ has been assumed in 
Refs.~\cite{Bres02,Bres03}, which corresponds to the filling of the 
single-particle levels $1s^22s$. In contrast, the  $1s^21p$ HO ordering, which 
has been assumed in Refs.~\cite{Guar02,Guar03,Guar03a}, results in an angular 
momentum $L=1$. The comparison of respective binding energies indicates that
the $L=1$ state has a lower energy than the $L=0$ one. In conclusion,  a more
general criterion to select the shell ordering needs to be specified.

The aim of this paper is to determine the excitation spectrum of a single $^3$He
atom bound to a $^4$He$_N$ cluster. The ordering of these single-particle
levels will be relevant to describe mixed clusters with a higher number of
$^3$He atoms. It is also worth noticing that the knowledge of the one-fermion
spectra in terms of the number of bosons is also  relevant to determine the
constants entering the rate equations which establish the formation chemical
process \cite{bruch02}, and thus the abundances in the production experiments. 
Our calculations are based in the DMC method, using an importance sampling 
wave function which carries out the orbital angular momentum of the $^3$He 
relative to the $^4$He$_N$ cluster. The DMC procedure is thus adapted so as 
to determine the lowest energy state of the subspace of orbital angular 
momentum $L$. In order to obtain the excited states within each subspace of 
angular momentum $L$ we use an optimized form of the upper bounds provided by 
the sum rules method.

The paper is organized as follows. In Section II we briefly review previous
investigations of $^4$He$_N$$^3$He clusters. In Section III we present a
detailed description of the method used to study the ground and excited states
of these clusters. Our results are presented and discussed in Section IV.
Some final comments are given in Section V.

\section{A survey of previous results on $^4$He$_N$$^3$He clusters}

A pure $^4$He$_{N+1}$ cluster is described by the Hamiltonian
\begin{equation}
\label{he4}
H = - \frac{\hbar^2}{2m_4} \sum_{i=1}^{N+1} \nabla_i^2 +
\sum_{i<j}^{N+1} V(r_{ij}) \, ,
\end{equation}
where $m_4$ is the mass of $^4$He, and $V(r)$ is the interaction potential.
Recent forms~\cite{HFD-B,LM2M2,TTY} of this interaction are quite similar and
we will use along the paper the one known as HFD-B~\cite{HFD-B} potential.

Given that the He-He interaction is a consequence of the interaction between
the electrons in the atoms, it is independent of the mass or the spin of
the nucleus. To convert the $(N+1)$-th atom into an  $^3$He atom, thus
dealing with the  cluster $^4$He$_N\,^3$He, corresponds to a simple change
in the Hamiltonian
\begin{equation}
\label{he3}
H = - \frac{\hbar^2}{2m_4} \sum_{i=1}^{N} \nabla_i^2
- \frac{\hbar^2}{2m_3} \nabla_{N+1}^2
+ \sum_{i<j}^{N+1} V(r_{ij}) \, ,
\end{equation}
where $m_3$ is the mass of $^3$He. The corresponding many-body problem is then
not much different from that of a pure $^4$He cluster. In the rest of the
paper we will use the subindex $F$ instead of $N+1$, to alleviate the notation.

Mixed $^4$He$_N$$^3$He clusters containing a single fermion have been
investigated using several theoretical methods. The first systematic study of
the excitation spectrum of the $^3$He atom was made by Dalfovo~\cite{Dalf89},
based on a zero-range density functional. The use of a non-local finite-range
density functional~\cite{Barr97} results in small quantitative differences,
related to the fact that finite-range functionals are more repulsive than
zero-range ones. As the size on the drop increases, the $^3$He atom is pushed
to the surface region, due to its large zero-point motion, and for large enough
clusters, the centrifugal term $L(L+1)/r^2$ entering the Schr\"odinger equation
for the $^3$He atom can be treated as a perturbation. Actually~\cite{nava04},
the general trend of the spectrum is a series of tight rotational bands on top
of radial excitations, related to the number of nodes in the radial wave
function. In the limit of very large $N$, the spectrum of the $^3$He atom
becomes independent of the quantum angular momentum, forming the analogous of
the two-dimensional Andreev states in bulk helium.

The so-called Lekner approximation was used in Ref.~\cite{Beli94}, where a VMC
calculation was performed, as well as in Ref.~\cite{Krot01}, based on the
Hypernetted-Chain method in its optimized version. Such approximation, used
by Lekner~\cite{Lekn70} to analyze the Andreev states in bulk liquid, assumes
that the pair correlations between $^3$He and $^4$He atoms are the same as those
between pairs of $^4$He atoms, so that the cluster $^4$He$_N$$^3$He can be
considered as a perturbation of the cluster $^4$He$_{N+1}$, the perturbation
being given by
\begin{equation}
\label{pert}
H_I =
\left( \frac{\hbar^2}{2m_4}- \frac{\hbar^2}{2m_3} \right) \nabla_{F}^2
\, .
\end{equation}
Further elaboration of the perturbation scheme results in a single-particle
Schr\"odinger equation  describing the $^3$He atom with an effective potential
given by
\begin{equation}
\label{v3}
V_3(r) =
\left( \frac{m_4}{m_3}-1\right) \tau_4(r)+
\frac{\hbar^2}{2 m_3} \frac{\nabla^2 \sqrt{\rho_4(r)}}{\sqrt {\rho_4 (r)}}
\, ,
\end{equation}
where  $\rho_4(r)$ and $\tau_4(r)$ are respectively the $^4$He particle and
kinetic energy densities, both defined in the unperturbed system with $N+1$ 
bosons. Notice that the Laplacian operator acting on the square root of 
$\rho_4$ produces a strongly attractive force peaked at the surface of the 
cluster.

In order to classify the resulting spectra, Krotscheck and Zillich~\cite{Krot01}
defined an effective wave number $k= \sqrt{L(L+1)}/R$ for each excitation
characterized by the orbital angular momentum $L$,  where $R$ is the equivalent 
hard sphere radius $R=\sqrt{5/3}r_{\rm rms}$, defined in terms of the root 
mean square radius $r_{\rm rms}$ of the droplet. By plotting all excitation 
energies as a function of $k$ for a large number of clusters, Krotscheck and 
Zillich found that all results fall reasonably well on a universal quadratic 
line, in nice agreement with the density-functional results.

Both sets of results may be approximately pictured as molecular rigid rotor, 
in which the $^4$He$_N\,^3$He cluster is viewed as a two body system, the 
cluster formed by the $N$ bosons plus the single fermion, tied by a spring 
with a rather large rigidity constant. Various angular momentum $L$ states are
associated to each vibrational-like state of the spring, obeying the law
\begin{equation}
\label{constants}
\delta E_L \approx \hbar^2 L(L+1)/2 {\cal I}
\equiv K L(L+1).
\end{equation}
This equation defines the rotational constant $K$ (units of energy) in terms
of the momentum of inertia $\cal I$, the later being proportional to $R^2$, 
where $R$ is the average distance of the fermion to the center-of-mass of the 
bosonic cluster.

In the case of light clusters ($N<40$) the calculations of Ref.~\cite{Krot01} 
find appreciable deviations from the universal behavior Eq. (\ref{constants}). 
Small $^4$He$_N$$^3$He clusters have been studied in Ref.~\cite{Bres00} based 
on the DMC method, and in Refs.~\cite{Guar02,Guar03,Guar03a} without the use 
of the Lekner approximation. To this respect it is worth stressing that the 
Lekner approximation is basically a {\em weak coupling} description of the 
interaction of a $^3$He atom with a $^4$He cluster, because it does not 
include the perturbation that the outer $^3$He atom should generate on the 
binding cluster. One may expect this picture to be satisfactory for large 
bosonic clusters, but inappropriate for small $N$. As a  consequence, the
interesting area to explore corresponds to the case of a small $^4$He
cluster, which may be appreciably modified by the $^3$He atom.

\section{Ground and excited states description}

To carry out the Diffusion Monte Carlo calculation requires an importance 
sampling or guiding wave function, which incorporates as much as possible the 
characteristics of the system to be described. In particular, due to the strong 
short-range repulsion of the atom-atom interaction, it is advisable to 
introduce at least two-body Jastrow correlations. Moreover, the guiding wave 
function must confine the system and, finally, it has to include the bosonic 
symmetry related to the $^4$He atoms and the desired quantum numbers for the 
$^3$He atom. In this Section we shall describe a variational wave function 
which will be used to calculate the lowest-energy states for a given value of 
the angular momentum $L$ of the system. Excited states corresponding to radial 
excitations will be estimated by means of sume-rule techniques.

\subsection{Ground state and angular momentum excitations}

In order to describe the system $^4$He$_N$$^3$He in a state where the $^3$He
atom is in an orbital angular momentum $L$ with respect to the $^4$He$_N$
system, a simple but nevertheless complete wave function is given by
\begin{eqnarray}
\label{importance_sampling}
\Psi({\mathbf r}_1, \dots, {\mathbf r}_N, {\mathbf r}_F)
&=&
 \Phi_{B}({\mathbf r}_1, \dots, {\mathbf r}_N) \\ \nonumber &&
\Phi_{M}({\mathbf r}_1,  \dots, {\mathbf r}_N, {\mathbf r}_F)
\Phi_{L}({\mathbf r}_F - {\mathbf R}_B) \, ,
\end{eqnarray}
where the subindexes $B$ and $F$ stand for bosons and fermions, respectively,
whereas $M$ refers to the mixed boson-fermion part of the wave function. The
bosonic coordinates run from 1 to $N$ and the coordinate of the fermion is
labelled by $F$. Finally, ${\mathbf R}_B$ represents the center-of-mass
coordinate of the bosonic subsystem. This model wave function includes
an internal bosonic part ($\Phi_B$) and the coupling of the fermion to
individual bosons ($\Phi_M$) as well as to the bosonic cluster ($\Phi_L$).

We have taken $\Phi_B$ and $\Phi_M$ to be of the Jastrow form
\begin{eqnarray}
\Phi_B({\mathbf r}_1 \dots {\mathbf r}_N) &=&
\prod_{i<j=1}^N {\mathrm e}^{f_B(r_{ij})} \label{boson} \\
\Phi_M({\mathbf r}_1 \dots {\mathbf r}_N, {\mathbf r}_F)
&=& \prod_{i=1}^N {\mathrm e}^{f_M(r_{iF})} \, , \label{mixed}
\end{eqnarray}
with
\begin{equation}
f_{B,M} (r) = - \frac{1}{2} \left(\frac{b_{B,M}}{r} \right)^{\nu} -p_{B,M}\, r 
\, .
\end{equation}
The two-body correlation terms include a short-range part, associated with the 
parameters $b_B$ and $b_M$, and a long-range confining part associated with 
the parameters $p_B$ and $p_M$. The short-range part is mainly related to the 
small-distance behavior of the relative two-body wave function. Consequently, 
the parameters $b_{B,M}$ and $\nu$ have been taken to be the same for all 
systems studied, and in our calculations they have been kept fixed to the values
$b_B = 2.95$~\AA, $b_M=2.90$~\AA~ and $\nu=5.2$, as obtained in our previous 
calculations for pure $^4$He and $^3$He clusters~\cite{Guar99,Guar00,Guar00a}, 
by direct minimization of the expectation value of the energy. On the other 
hand, the long-range confining parameters $p_{B,M}$ have been determined by 
means of the VMC method.

In the absence of the last term $\Phi_L$ of the importance sampling wave 
function (\ref{importance_sampling}) we would describe a state of null angular 
momentum, explicitly translational invariant and including the bosonic symmetry 
of the indistinguishable $^4$He atoms. The role of the last term $\Phi_L$ of  
Eq.~(\ref{importance_sampling}), describing the  motion of the fermion, is to 
determine  the value of the orbital angular momentum $L$. It has been taken as 
a long-range wave function depending on the relative coordinate of the fermion 
with respect to the center-of-mass of the bosons, 
${\bf r} = {\bf r}_F - {\bf R}_B$. The explicit form used is the harmonic 
polynomial
\begin{equation}
\label{fermion}
\Phi_{L}({\mathbf r}) = r^L P_L(\cos \theta_r) \, .
\end{equation}
This function is particularly simple and corresponds to a state with orbital
angular momentum $L$ and null third component. One could have taken a more
sophisticated form, by putting a radial dependence different from the simple 
$r^L$, but it is reasonable to expect  the DMC algorithm to be able to improve
this simple and computationally convenient form. By making  this part of the
trial wave function to depend on the relative distance of the fermion to the
center-of-mass of the bosons the translational invariance of the importance 
sampling wave function is not spoiled. Notice that if we had considered the 
function to depend on the distance of the fermion to the center-of-mass of the 
full system, the only difference would have been a trivial scale factor.

With the structure of the importance sampling wave functions one may describe
the lowest-energy states for each angular momentum $L$. It should be mentioned
that apart from the  case $L=0$ for which $\Phi_L=1$, all other cases 
correspond to functions with a nodal surface, with nodes depending only on the 
angular variables. This fact has to be taken into account when using the DMC 
algorithm.

\subsection{Radial excitations}

When considering a subspace of angular momentum $L$ the DMC algorithm gives
only the energy of the ground state of that subspace, and there is no
information about the excited states of the same angular momentum. A way to
have an estimate, actually an upper bound, of the first excited state is to use
the sum rules method~\cite{Guar01}.

Consider the exact ground state for a given angular momentum $L$, represented
here by $\Psi_{0L}$, and the full set of eigenstates of this subspace ordered 
by increasing energy and represented by $\{ \Psi_{nL},E_{n,L}\}, n=0, 1 \dots$. 
Let $Q({\bf R})$ be an arbitrary Hermitian operator which may depend on all 
atomic coordinates, which is  assumed to be  scalar under rotations, i.e.,
to commute with ${\bf L}$ and ${\bf S}$. Let us consider the sum rule of
order $p$
\begin{equation}
\label{sum_rules}
M^{(p)}_L [Q] = \sum_{(n,\ell)\neq (0,L)} (E_{n\ell} - E_{0L})^p
 |\langle\Psi_{n\ell}|Q|\Psi_{0L}\rangle|^2 \, ,
\end{equation}
where the sum extends to all eigenstates of the Hamiltonian but the lowest 
energy state of angular momentum $L$. This is important because in order to 
obtain easily computable properties  it will convenient  to use the 
completeness relation. Because of the assumed properties of $Q$ only states 
with  angular momentum $\ell=L$ will contribute to the sum. The $p=1$
rule fulfills the property
\begin{eqnarray}
M_L^{(1)}[Q] & \equiv &
\sum_{n\neq 0} (E_{nL}-E_{0L}) |\langle\Psi_{nL}|Q|\Psi_{0L}\rangle|^2 
\nonumber \\
& \geq & (E_{1L}-E_{0L}) M_L^{(0)}[Q] \, ,
\end{eqnarray}
from which one obtains an upper bound to the energy of the first excited state
of the subspace $L$
\begin{equation}
\label{upper_bound}
E_{1L} - E_{0L} \leq\frac{M_L^{(1)}[Q]}{M_L^{(0)}[Q]} \, .
\end{equation}

The evaluation of the sum rules is simpler than  seem, because of the
relations
\begin{eqnarray}
\label{eme0}
M_L^{(0)}[Q] &=& \langle\Psi_{0L} |  Q^2  | \Psi_{0L} \rangle
- |\langle\Psi_{0L} |  Q  | \Psi_{0L} \rangle |^2 \\
\label{eme1}
M_L^{(1)}[Q] &=& \frac{1}{2}\langle\Psi_{0L} |  [Q,[H,Q]] | \Psi_{0L} \rangle 
\, .
\end{eqnarray}
The double commutator may be simplified for a general Hamiltonian of the form 
given in Eq.~(\ref{he3}), obtaining
\begin{eqnarray}
\label{simple_m1}
M_L^{(1)} &=& \frac{\hbar^2}{2m_4} \langle \Psi_{0L} | \sum_{i=1}^N | 
\nabla_i Q |^2 |\Psi_{0L} \rangle \\ \nonumber
&+& \frac{\hbar^2}{2m_3}  \langle \Psi_{0L} | |\nabla_F Q |^2 | \Psi_{0L} 
\rangle \, .
\end{eqnarray}
Note that to compute these expressions one only requires knowledge of the
 ground state wave function of the angular momentum $L$ subspace.

This method was used in Ref.~\cite{Guar01} to obtain upper bounds to the
first $L=0$ excitation, as well as to the low-lying even-L states. Given that 
we are obtaining the $L\neq 0$ excitations directly from the DMC procedure, 
the sum rules method will be used here to obtain the energies of the first 
excited states in each $L$-subspace. In the Appendix we use the sum rule 
method to also estimate $L\neq 0$ excitations based in the knowledge of 
$\Psi_{00}$ by relaxing the scalar character of $Q$ as an alternative to the 
direct DMC calculations.

The upper bound given by Eq.~(\ref{upper_bound}) is a functional of the
operator $Q$, so it may be variationally optimized by equating to zero its
functional derivative with respect to $Q$. Unrestricted minimization will give
rise to the unpractical relation $Q|\Psi_{0L}\rangle = |\Psi_{1L}\rangle$, its
solution being equivalent to the solution of the many-body Schr\"odinger
equation for the excited state. An alternative is to optimize the operator
inside a restricted subspace, which is the approach followed by
Chin and Krotscheck~\cite{Chin90,Chin92} and is closely related to the
procedure of Krisna and Whaley~\cite{Kris90}.

Here we have followed a simpler procedure, based in the linear expansion of the
operator in a basis of easily computable operators. To determine the basis we
have assumed a single-particle-like form for the operator, by considering that
it depends only on ${\bf r}_F - {\bf R}_B$, i.e., on the coordinate of the
fermion referred to the bosonic center-of-mass
\begin{equation}
\label{Qfunction}
Q = Q({\bf r}_F - {\bf R}_B).
\end{equation}
This simple form preserves the translational invariance and does not spoil the
boson symmetry of the $^4$He subsystem. For this type of general operator
Eq.~(\ref{simple_m1}) can be further simplified, since the following relation
$$
\nabla_i Q({\bf r}_F - {\bf R}_B) = - \frac {1}{N} \nabla_F Q ({\bf r}_F -
{\bf R}_B)
$$
holds. The resulting sum rule $M_1$ becomes
\begin{equation}
\label{simplerr_m1}
M_1 = \frac{\hbar^2}{2 \mu} \langle \Psi_{0L} | \, |\nabla_F Q|^2 | \Psi_{0L} 
\rangle \, ,
\end{equation}
where $\mu= (N m_4  m_3) / (Nm_4 + m_3)$ is the reduced mass of the $^3$He 
atom and the $^4$He cluster.

In the calculations to be described below  the monopole operator has been
optimized by using a simple functional form depending on few parameters,
\begin{equation} 
Q ({\bf r}_F - {\bf R}_B)  = \sum_{m=1}^5  C_m q_m ({\bf r}_F - {\bf R}_B) \, ,
\end{equation}
with
\begin{equation}
\label{q_operator}
q_m({\bf R}) = |{\mathbf r}_F -{\mathbf R}_B|^m \, .
\end{equation}

The minimization of the  upper bound of Eq.~(\ref{upper_bound}) with respect to
$C_m$, for angular momentum $L$, gives rise to a generalized eigenvalue problem,
\begin{equation}
\label{valorpropio}
{\cal M}^{(1)}_{mn} C_n = {\cal E} {\cal M}^{(0)}_{mn}C_n \, ,
\end{equation}
with a Hamiltonian-like matrix
\begin{equation}
\label{h_matrix}
{\cal M}^{1}_{mn} = \frac{\hbar^2}{2\mu}
\langle  \Psi_{0L} | \nabla_F q _m  \nabla_F q_n | \Psi_{0L} \rangle
\end{equation}
and a normalization matrix
\begin{eqnarray}
\label{m_matrix}
{\cal M}^0_{mn} &=& \langle \Psi_{0L} | q_m q_n | \Psi_{0L} \rangle \\
\nonumber
& - &
\langle \Psi_{0L} | q_m  | \Psi_{0L} \rangle
\langle \Psi_{0 L}|  q_n | \Psi_{0L} \rangle.
\end{eqnarray}
The lowest eigenvalue of Eq.~(\ref{valorpropio}) provides an optimized upper 
bound. By inserting Eq.~(\ref{q_operator}) into Eqs.~(\ref{h_matrix}) and 
(\ref{m_matrix}) the matrices are further simplified to
\begin{equation}
\label{H_matrix_simp}
{\cal M}^{(1)}_{mn} = \frac{\hbar^2}{2\mu}
m n
\langle  \Psi_{0L} | \, |{\bf r}_F - {\bf R}_B|^{m+n-2} | \Psi_{0L} \rangle
\end{equation}
and
\begin{eqnarray}
\label{m_matrix_simp}
{\cal M}^{(0)}_{mn} &=& \langle \Psi_{0L} | \,
 |{\bf r}_F - {\bf R}_B|^{m+n} | \Psi_{0L} \rangle
\\ \nonumber
&-&
\langle \Psi_{0L} | \, |{\bf r}_F - {\bf R}_B|^{m}  | \Psi_{0L} \rangle
\langle \Psi_{0L} |  \, |{\bf r}_F - {\bf R}_B|^{n} | \Psi_{0L} \rangle.
\end{eqnarray}
It is worth keeping in mind that in our DMC calculations the matrix elements 
${\cal M}^{(p)}_{mn}$ are based in  a mixed estimate, so the strict variational 
character of Eq.~(\ref{upper_bound}) may be lost.

Some insight on the structure of these excitation may be drawn by considering 
the leading $m=n=1$ terms of both the Hamiltonian and norm matrices. For this 
one-dimensional subspace the upper bound to the radial excitation energy is 
given by
\begin{eqnarray}
\label{rad_excit_0}
&& E_{1L} - E_{0L} \leq \\ \nonumber
&& \frac{\hbar^2}{2\mu} \frac{1}{\langle 0L|\, |{\bf r}_F - {\bf R}_B|^{2}
 | 0L\rangle
 - |\langle 0L|\, |{\bf r}_F - {\bf R}_B|\, | 0L \rangle  |^2 } \, .
 \end{eqnarray}
The size of this bound depends on the difference between the mean square
radius and the squared mean radius of the fermion with respect to the boson
center-of-mass. This difference will be small if the distribution of the
fermion with respect to the boson center-of-mass is sharply peaked near a
given value say $R_0$, corresponding to a rather rigid spring. Then the
denominator will be small and the radial excitation will have large energy.

The excited states considered in this section  result from excitations related 
to the distance between the fermion and the center-of-mass of the cluster of 
bosons. The underlying optimal wave function (not determined, however, from 
the DMC procedure) should have at least a node along this radial coordinate, 
in order to be orthogonal to the angular momentum $L$ ground state, and may be 
properly termed as a {\em radial excitation}.

\subsection{Computational details}

The DMC algorithm~\cite{Reyn82,Mosk82} is nowadays a well-known and used
technology. It is based in integrating the imaginary-time Schr\"odinger
equation for an auxiliary function
$f({\bf R} ,t) = \Phi_{\rm var}({\bf R}) \Psi({\bf R}, t)$ which is the
product of a trial wave function $\Phi_{\rm var}$ and the true ground-state
wave function $\Psi({\bf R}, t)$. The solution is given in terms
of an approximate small-time Green funtion $G({\bf R},{\bf R}', \tau)$
\begin{equation}
f({\bf R'},t+\tau) =\int d{\bf R} G({\bf R}',{\bf R}, \tau)f({\bf R},t)
\end{equation}
by means of a series of small time steps $\tau$. We have used the $O(\tau^3)$
approximate Green function~\cite{Vrbi86,Chin90a} which provides an $O(\tau^2)$
approximation for the energy. In our calculations we have used a set of 1000 
walkers, on the average, a value of $\tau= 0.0002\, {\rm K}^{-1}$, 8000 
iterations to settle down the system and 80000 iterations to compute the 
averages. For $N \geq 40$ we have used a smaller time step
$\tau = 0.00015\, {\rm K}^{-1}$ and the number of iterations has been doubled.

The wave function is not definite positive when $L \neq 0$, and this leads to 
the known and irritating {\em sign problem} of the DMC algorithm. We have used 
the so-called {\em fixed node approximation}. The auxiliary function 
$f({\bf R}, t)$ will remain positive if both functions $\Phi_{\rm var}$ and 
$\Psi$ have (at any time) the same nodal surfaces. The fixed node approximation 
consists in killing  any walker which attempts to cross a nodal surface.  
It has been shown~\cite{Cepe81,Reyn82,Mosk82} that this procedure leads to an 
upper bound to the  ground state.

\section{Energetics of $^4$He$_N$$^3$He clusters}

We have first performed a VMC calculation to determine the free parameters
$p_B$ and $p_M$ of the importance sampling wave function. As mentioned in the
previous Section, the parameters $b_B$, $b_M$ and $\nu$ are fixed by the
atom-atom interaction at very short distances, and do not depend on the size
of the cluster. On the contrary, the long-range parameters $p_{B,M}$ are very
sensitive to the number of bosons. Their values, determined by minimizing the
ground state expectation value of the Hamiltonian, are reported in
Table~\ref{params}.

\begin{table}[h!]
\caption{Optimal values of the parameters (in \AA$^{-1}$)
 for the variational description of
the system $^3$He$^4$He$_N$, as a function of the number of bosons $N$.
The last two columns contain the ground state VMC and DMC energies (in K).
Figures in parenthesis are the standard deviations of the Monte Carlo
calculations.}
\label{params}
\begin{tabular}{rrrrr}
\hline
$N$ & $p_B$  &  $p_M$ & $E_{\mathrm VMC}$ & $E_{\mathrm DMC}$ \\
\hline
 5  & 0.1250 & 0.0900 &     $ -1.59(2)$      & $ -1.862(4)$\\
10  & 0.0672 & 0.0395 &     $ -8.99(6)$      & $ -9.763(9)$\\
15  & 0.0491 & 0.0321 &     $-19.17(13)$      & $-21.39(2)$\\
20  & 0.0365 & 0.0226 &     $-31.09(12)$      & $-35.42(3)$\\
25  & 0.0298 & 0.0187 &     $-44.51(16)$      & $-51.20(3)$\\
30  & 0.0253 & 0.0160 &     $-59.2(2)$      & $-68.03(5)$\\
35  & 0.0220 & 0.0140 &     $-74.5(3)$      & $-86.13(6)$\\
40  & 0.0195 & 0.0125 &     $-90.5(3)$      & $-104.90(9)$\\
45  & 0.0176 & 0.0113 &     $-106.1(3)$     & $-124.63(5)$ \\
50  & 0.0160 & 0.0104 &     $-122.8(5)$     & $-144.67(5)$ \\
\hline
\end{tabular}
\end{table}

One can see that the values of these parameters decrease when the number of 
bosons increase, as it should correspond to a drop, its size growing with the 
number of constituents. Notice that, for any given number of bosons, the 
parameter $p_M$, controlling the boson-fermion distance, is significantly 
smaller than the parameter $p_B$ controlling the boson-boson distance. This 
reflects the fact that the light particle stands near the surface of the 
bosonic drop.

\begin{figure}[h!]
\caption{Raw DMC results (boxes with error bars) and the least-squares fits.
From top to bottom, the figures show the dissociation energy and the
excitation energies for $L=1-4$, all measured in K, as a function of the
number of bosons $N$. The dashed line in the lower four figures is the
smoothed dissociation limit. }
\label{fits}
\includegraphics[width=6cm]{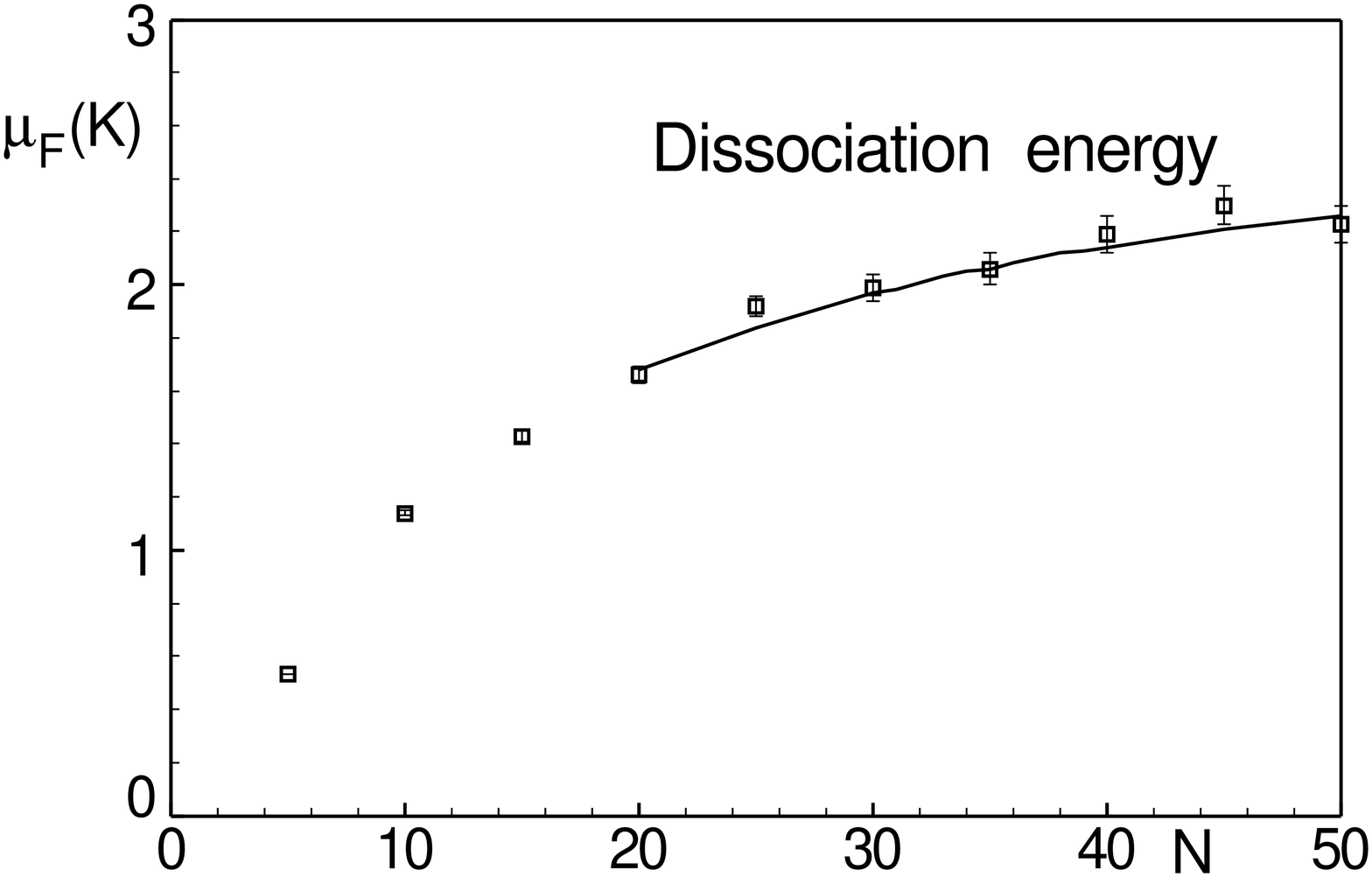}
\includegraphics[width=6cm]{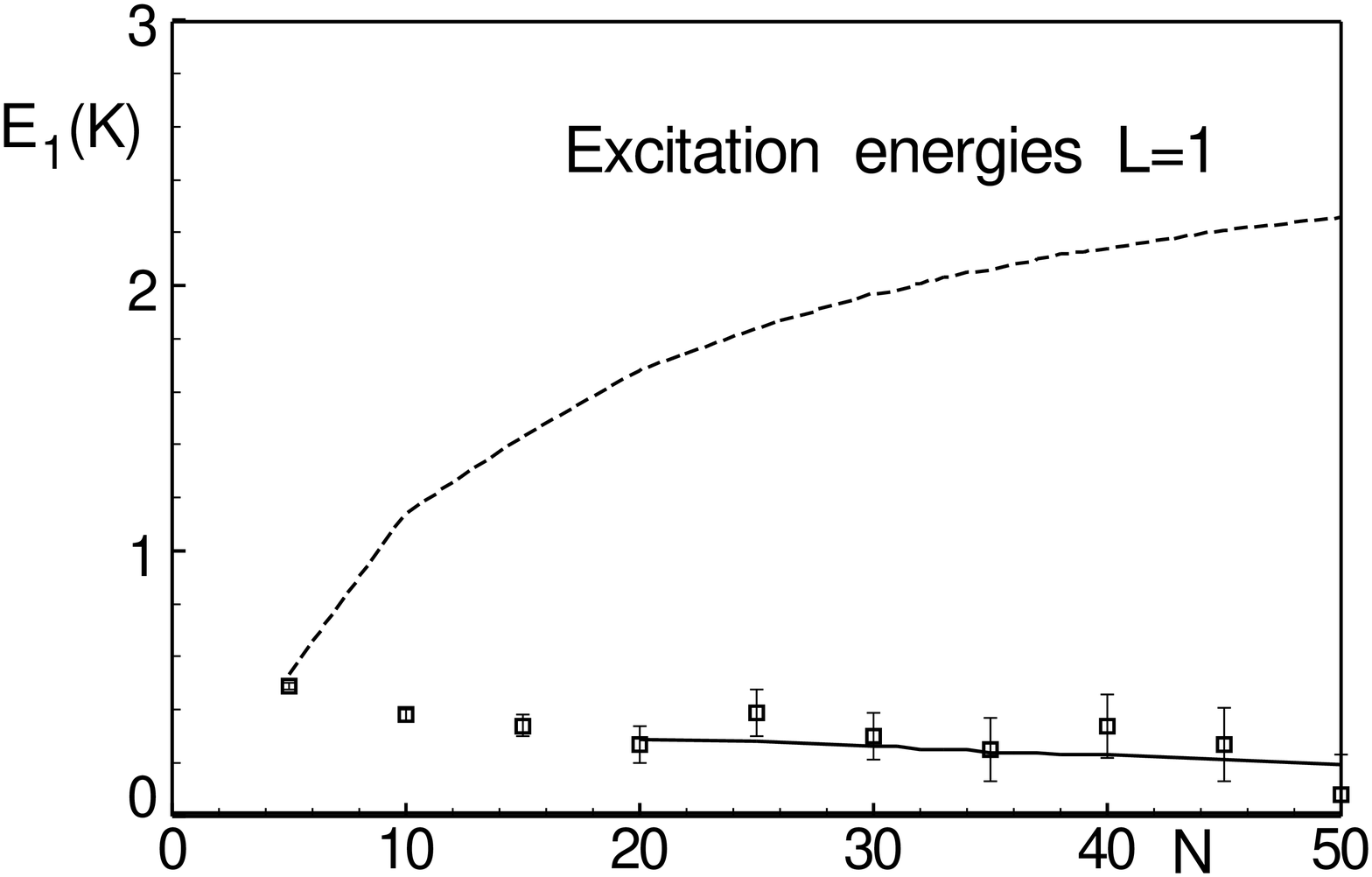}
\includegraphics[width=6cm]{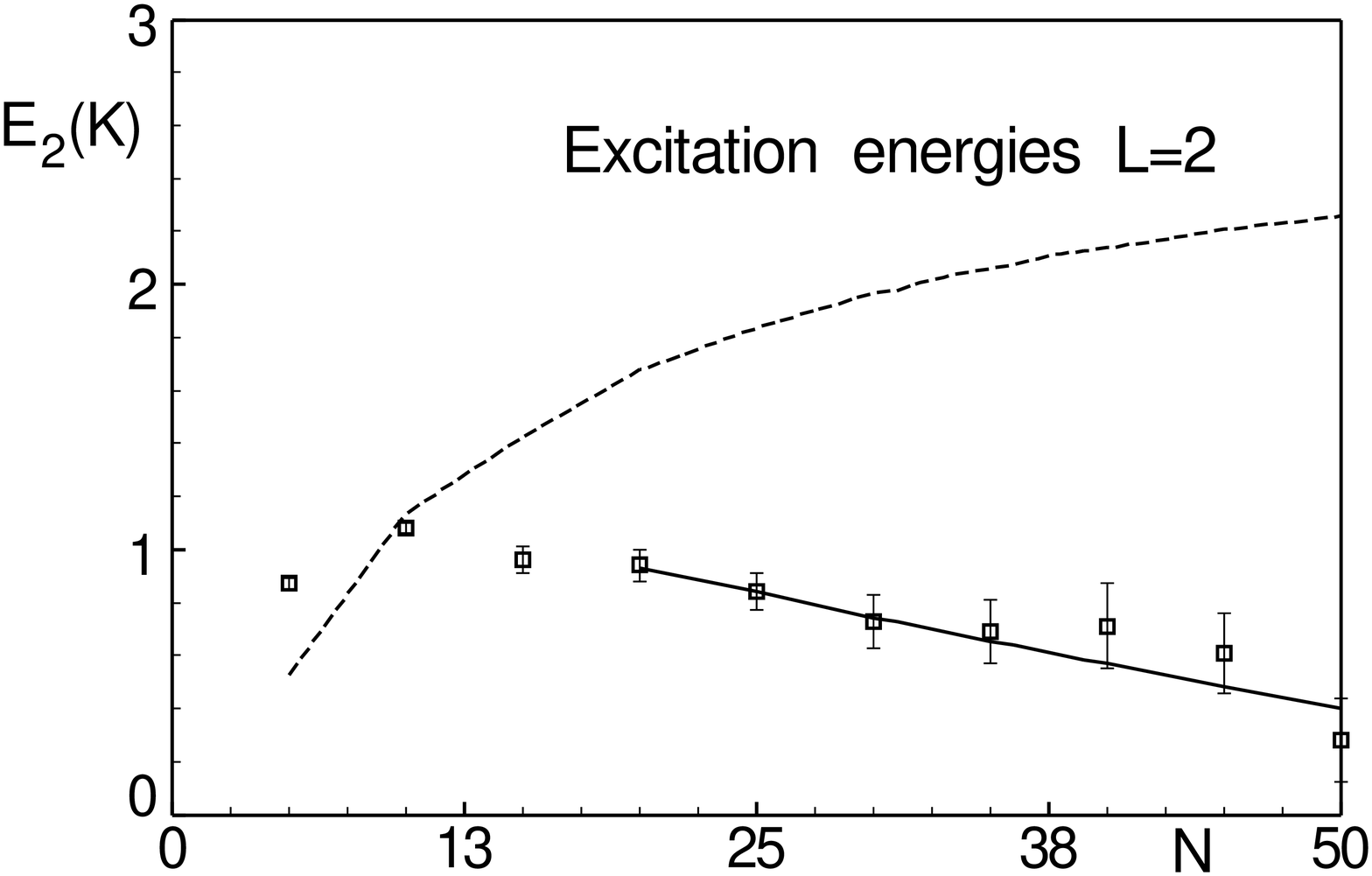}
\includegraphics[width=6cm]{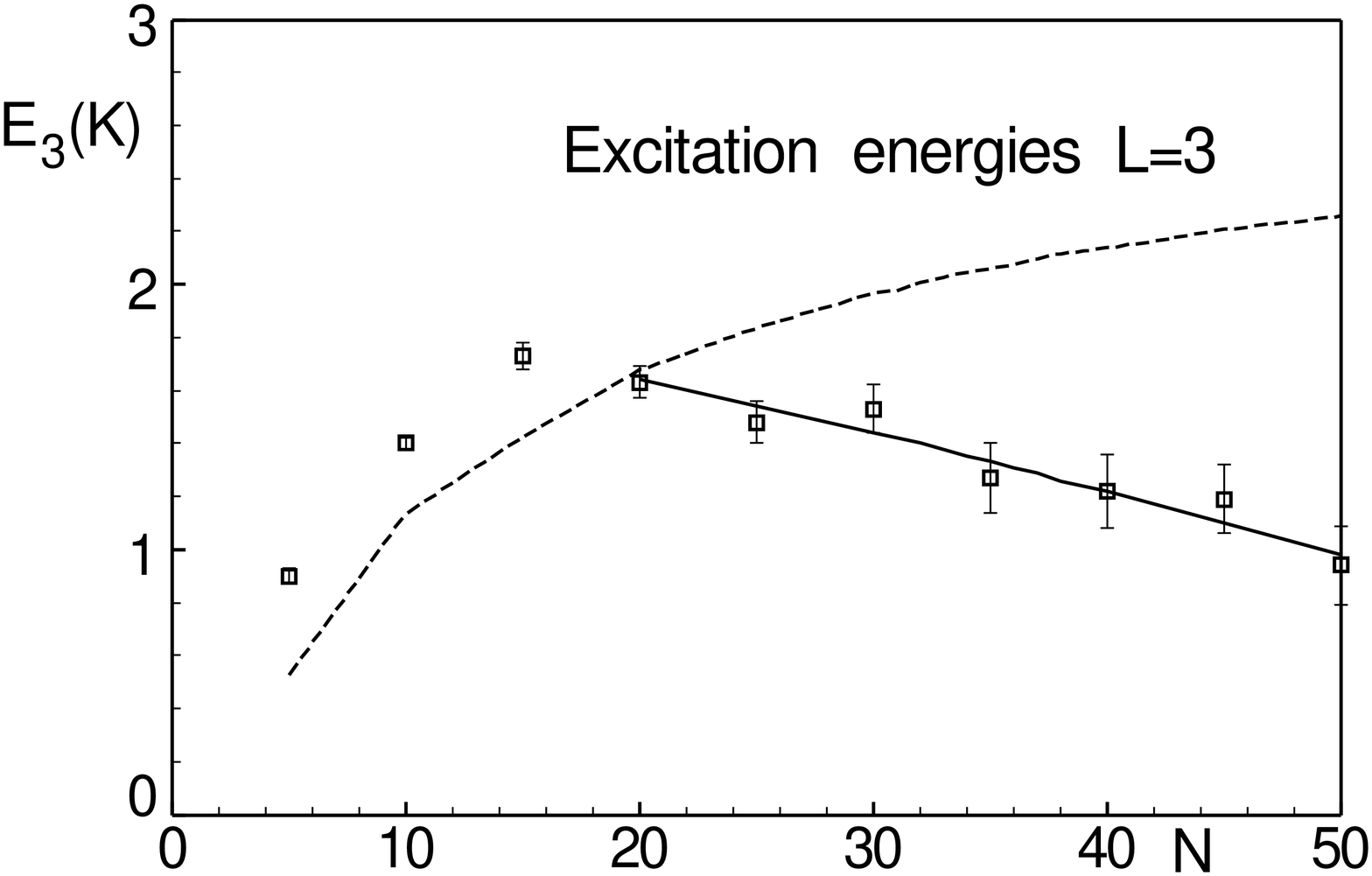}
\includegraphics[width=6cm]{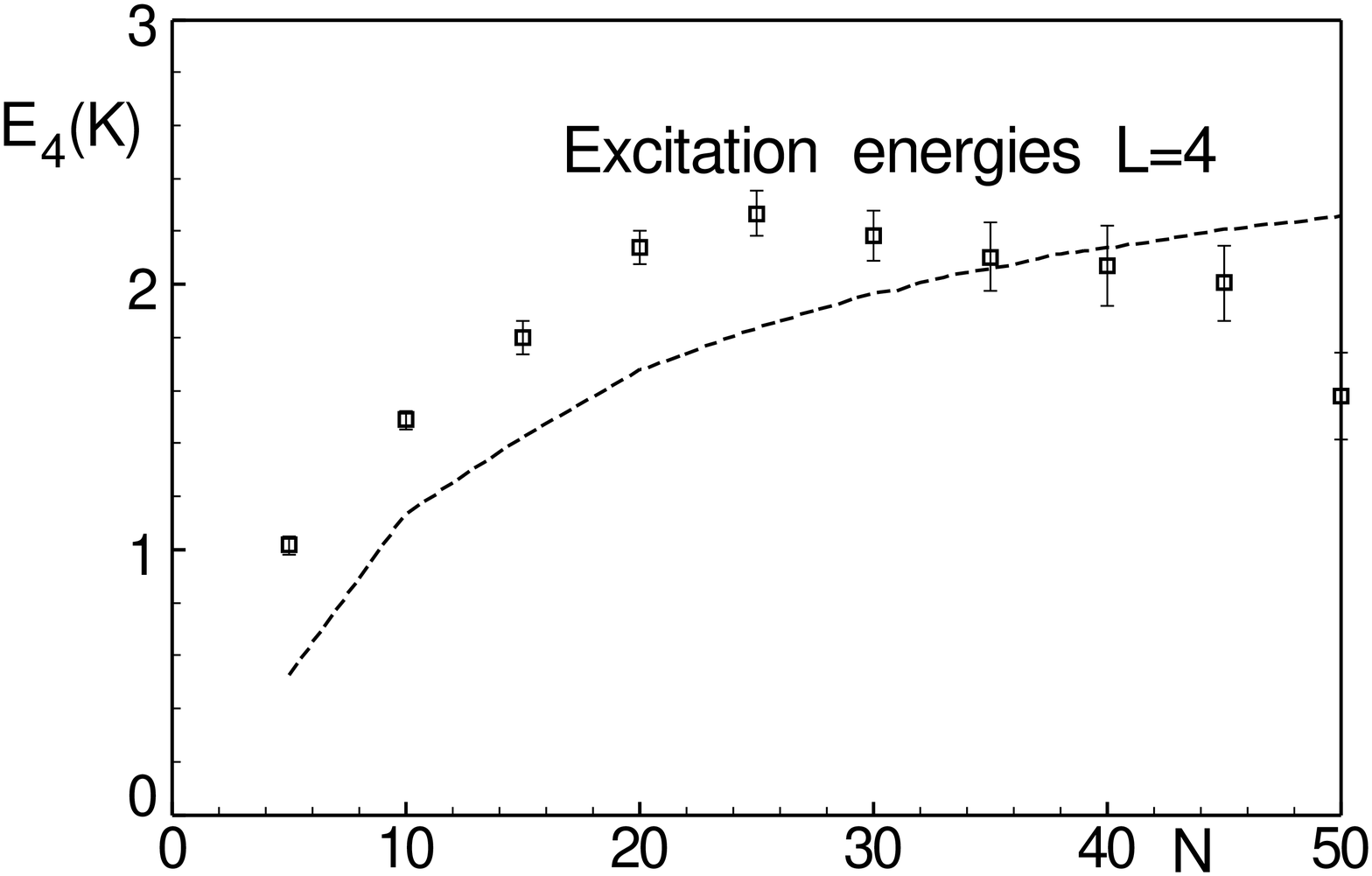}
\end{figure}

In Table~\ref{params} the ground state energies obtained within the VMC 
calculation as well as with the improved DMC method are also displayed. The 
later are significantly lower than the former, thus revealing that the
variational trial functions are not of high quality. They could be improved by 
adding either medium-range terms to the two-body Jastrow correlation or 
three-body Jastrow correlations. Notice that improving the importance sampling
wave function will not affect the DMC results of Table~\ref{params} except for 
the statistical error. Nevertheless, it could lead to better upper bounds for
the $L\neq 0$ energies.

The statistical errors of the DMC energies grow steadily with the number of
bosons. As a consequence, the determination of the excitation energies becomes
less and less accurate  for large $N$. The excitation energies $E_L$ are 
obtained as the difference of two independent calculations, one for $L=0$ 
($E_{00}$) and the other for the desired value of the angular momentum 
($E_{0L}$). These two energies have very close values, thus magnifying the 
statistical error of their difference. As a consequence, the direct plot of 
the excitation energies will show fluctuations. In order to control them we 
have fitted all cases between $N=30$ and $N=50$ with a liquid-drop like formula,
i.e. a third-order polynomial fit in terms of the variable $N^{1/3}$.

The $^3$He chemical potential, or $^3$He dissociation energy, is defined as
\begin{equation}
\label{dissociation}
\mu_F = E_{00}(^4{\rm He}_N) - E_{00}(^4{\rm He}_N \, ^3{\rm He}) \, .
\end{equation}
It corresponds to the energy required to eject the $^3$He atom from the
mixed system in its ground state ($L=0$). According to the definition, $\mu_F$
is a positive quantity and its value is relevant because the states whose
excitation energy is above it are not bound. To control the statistical
fluctuations in $\mu_F$ we have again fitted the raw differences with a
liquid-drop formula. In Fig.~\ref{fits} are plotted the raw DMC results for
the chemical potentials and the excitation energies, as well as their
respective fits as described above. For $L=4$ there were too few points
to carry out that fit, and we have only plotted the raw DMC results.

The corresponding values for bound levels are displayed in Table~\ref{spectrum}
for excited states with $L=1$ to $L=4$ and for systems with different number 
of bosons. The total energy of the ground state ($E_{00}$) has been already 
quoted in the last column of Table~\ref{params}. The excitation energies 
displayed in Table~\ref{spectrum} are the raw DMC differences for systems with 
$N<20$ and the results of the least squares fit otherwise.

\begin{table}[h!]
\caption{Excitation energies $E_L$, in K, of $^4$He$_N$$^3$He clusters
for $L=1-4$ and $N=5-50$ in steps of 5 atoms.
The values quoted are the result of the least squares fit described
above for $N\geq 20$ and raw DMC results
for $N<20$, with the exception of the $L=4$ column which contains the raw DMC
results. The last column displays the $^3$He dissociation limit.}
\label{spectrum}
\begin{tabular}{rllllll}
\hline
$N$ & $\Delta E_1$ & $\Delta E_2$ & $\Delta E_3$ &
$\Delta E_4$ & $\mu_F$ \\
\hline
5   &  0.48(1) &      &           &          & 0.53 \\
10  &  0.37(2) & 1.08(2) &           &          & 1.14 \\
15  &  0.33(4) & 0.96(5) &           &          & 1.43 \\
20  &  0.28 & 0.93 & 1.64      &          & 1.68 \\
25  &  0.27 & 0.84 & 1.54      &          & 1.85 \\
30  &  0.25 & 0.74 & 1.44      & 2.18(10) & 1.97 \\
35  &  0.23 & 0.65 & 1.33      & 2.10(12) & 2.05 \\
40  &  0.22 & 0.57 &  1.22     & 2.07(15) & 2.14 \\
45  &  0.20 & 0.48 &  1.10     & 2.01(14) & 2.21 \\
50  &  0.18 & 0.40 &  0.98     & 1.58(16) & 2.26 \\
  \hline
\end{tabular}
\end{table}

\begin{figure}[h!]
\caption{Radial excitations (in K) on top of $L=0$ to $L=2$ lowest levels.
The squares are the results obtained directly from the DMC calculation, the
continuous line is a guiding line for these excitations and the dashed line
is the dissociation limit.}
\label{radialexcit}
\includegraphics[width=6cm]{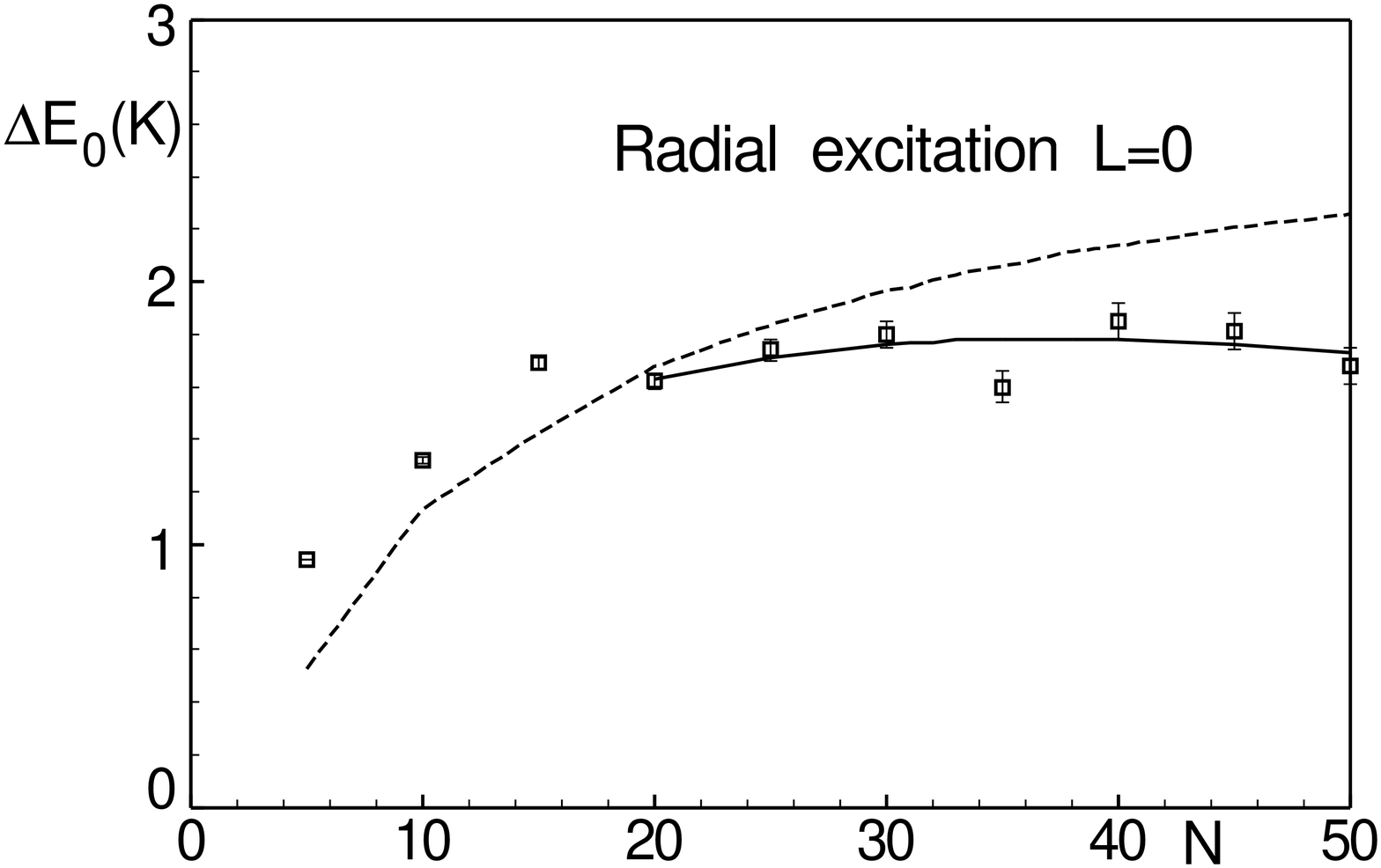}
\includegraphics[width=6cm]{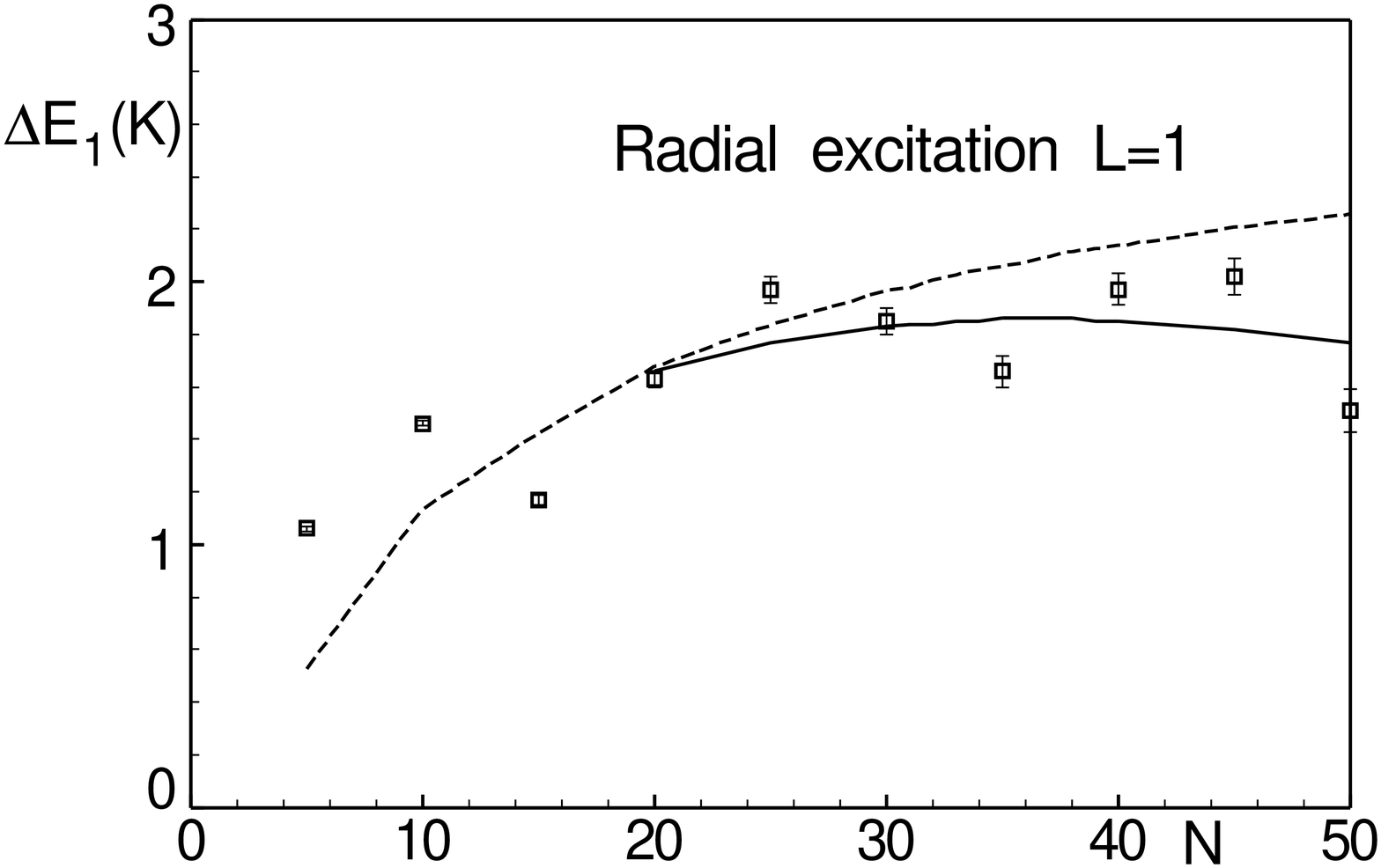}
\includegraphics[width=6cm]{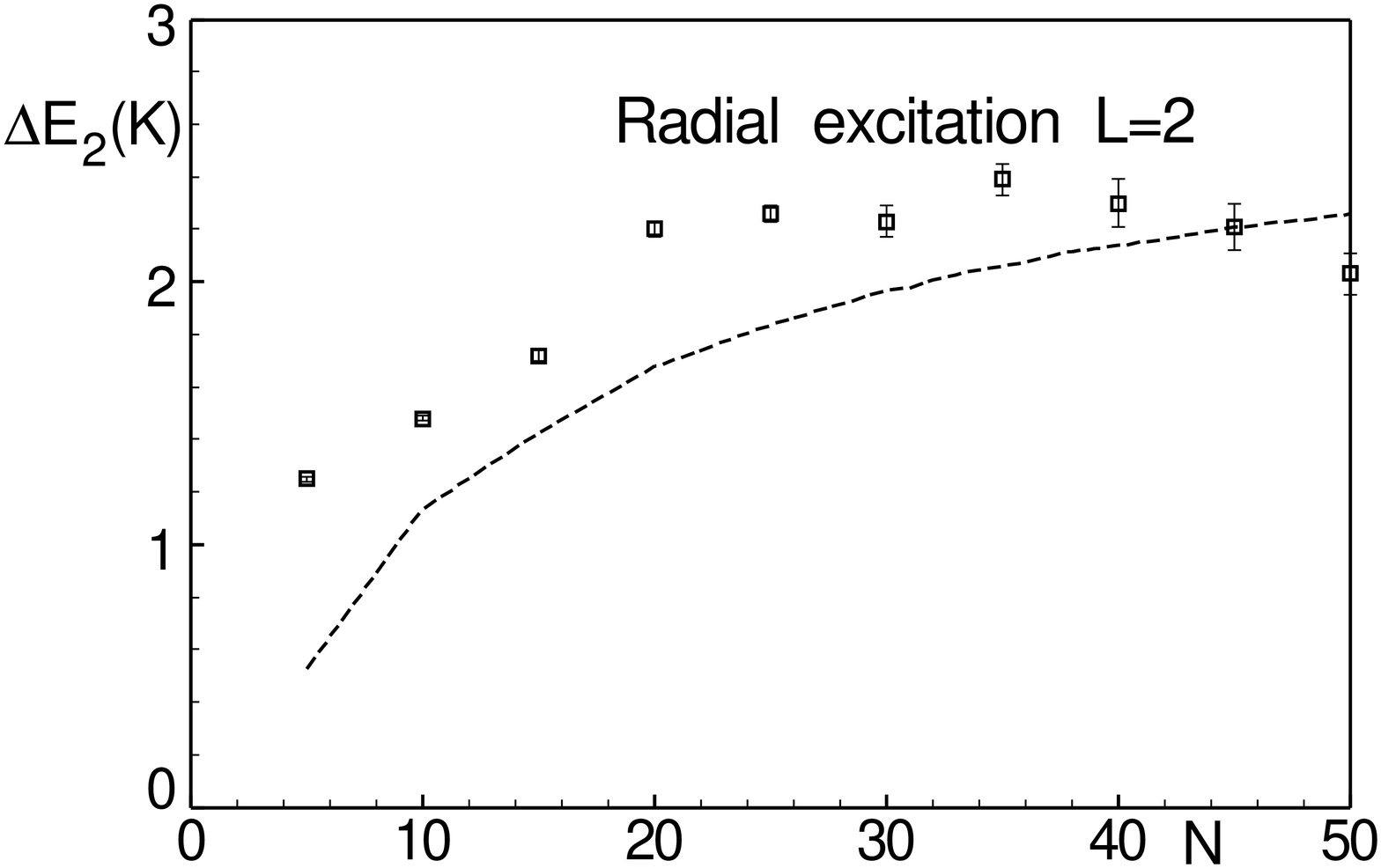}
\end{figure}

The excitation energies of radially excited levels obtained with the sum rules
method described above are shown in Fig.~\ref{radialexcit} for $L=0, 1, 2$. 
These levels are close to the dissociation limit, above it for $N<20$ and below 
afterwards, for $L=0,1$. The radial excitation for $L=2$ is always above the 
dissociation limit, with the exception of $N$ near $50$, which signals the 
threshold for the binding of this level. One should keep in mind that the upper 
bound character of the excitation energies, as expressed by 
Eq.~(\ref{upper_bound}), is not strictly satisfied in the present calculations 
because the radial excitation energies have been computed by means of 
(approximate) mixed matrix elements; therefore, some of the referred levels 
may not be  bound in reality. For $L>2$ the radial excitations are clearly 
unbound up to $N=50$.

\begin{figure}[h!]
\caption{Excitation energies of the $^3$He bound excited levels in a mixed
droplet as a function of the number of $^4$He atoms. The levels are ordered by
the value $L$ of the angular momentum, written in the right side. Index
$0^*$ refers to the to the radial excitations of the ground state
and corresponds in the figure to dashed lines. The dotted line is a
guiding line joining the chemical potential $\mu_F$.}
\label{spectra}
\includegraphics[width=6.8cm,angle=-90]{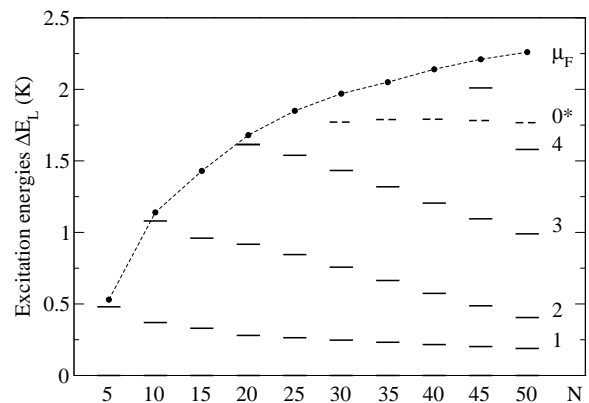}
\end{figure}

The bound level spectra resulting from our calculations are collected up in
Fig.~\ref{spectra}. The bound levels are grouped by the value $L$ of the
angular momentum,  indicated in the right side of the figure, with
the symbol $0^*$ signaling the radial excitations of the ground state.

As we have mentioned in Section II, one could expect these spectra to look
like a series of rotational bands, with the excitation energies roughly
proportional to the value $L(L+1)$, where $L$ is the angular momentum of the
excited level. That is to say that the quantities
\begin{equation}
\label{rotor}
K = \frac{E_{0L} - E_{00}}{L(L+1)} \, ,L\neq 0\, ,
\end{equation}
which correspond to the rotational constants ({\it cf.} Eq.~\ref{constants}), 
should be expected to depend only on $N$, and not on the angular momentum $L$. 
This is indeed the case, as shown in Table~\ref{rotational_constants}. The 
values of the rotational constants smoothly decreases as $N$ increases, as 
expected because of the dominant dependence on  
$\langle 1/|{\bf r}_F- {\bf R}_B|^2\rangle$.

\begin{table}[h!]
\caption{Rotational constants, in K, as defined in
Eq.~(\ref{rotor}), for the bound levels with $L=1-4$.}
\label{rotational_constants}
\begin{tabular}{rrrrrr}
\hline
$N$ & $L=1$ & $2$ & $3$ & $4$  \\
\hline
5   &  0.24 &      &           &       \\
10  &  0.19 & 0.18 &           &       \\
15  &  0.17 & 0.16 &           &       \\
20  &  0.14 & 0.16 &  0.14     &       \\
25  &  0.14 & 0.14 &  0.13     &       \\
30  &  0.13 & 0.12 &  0.12     &       \\
35  &  0.12 & 0.11 &  0.11     & 0.11  \\
40  &  0.11 & 0.10 &  0.10     & 0.10  \\
45  &  0.10 & 0.08 &  0.09     & 0.10  \\
50  &  0.09 & 0.07 &  0.08     & 0.08  \\
  \hline
\end{tabular}
\end{table}

\section{Structure of $^4$He$_N$$^3$He clusters}

\begin{figure}[h!]
\caption{The angular-averaged density distributions of the fermion with
respect to the center-of-mass of the system $^4{\rm He}_{N}\,^3{\rm He}$ for
several values of the number of bosons $N$ and the fermionic orbital angular
momentum $L$. The boson density distribution with respect to the
center-of-mass appearing in the plot has been arbitrarily divided by 20
to appreciate the fermion distributions in the figures.}
\label{distributions}
\includegraphics[width=4.25cm]{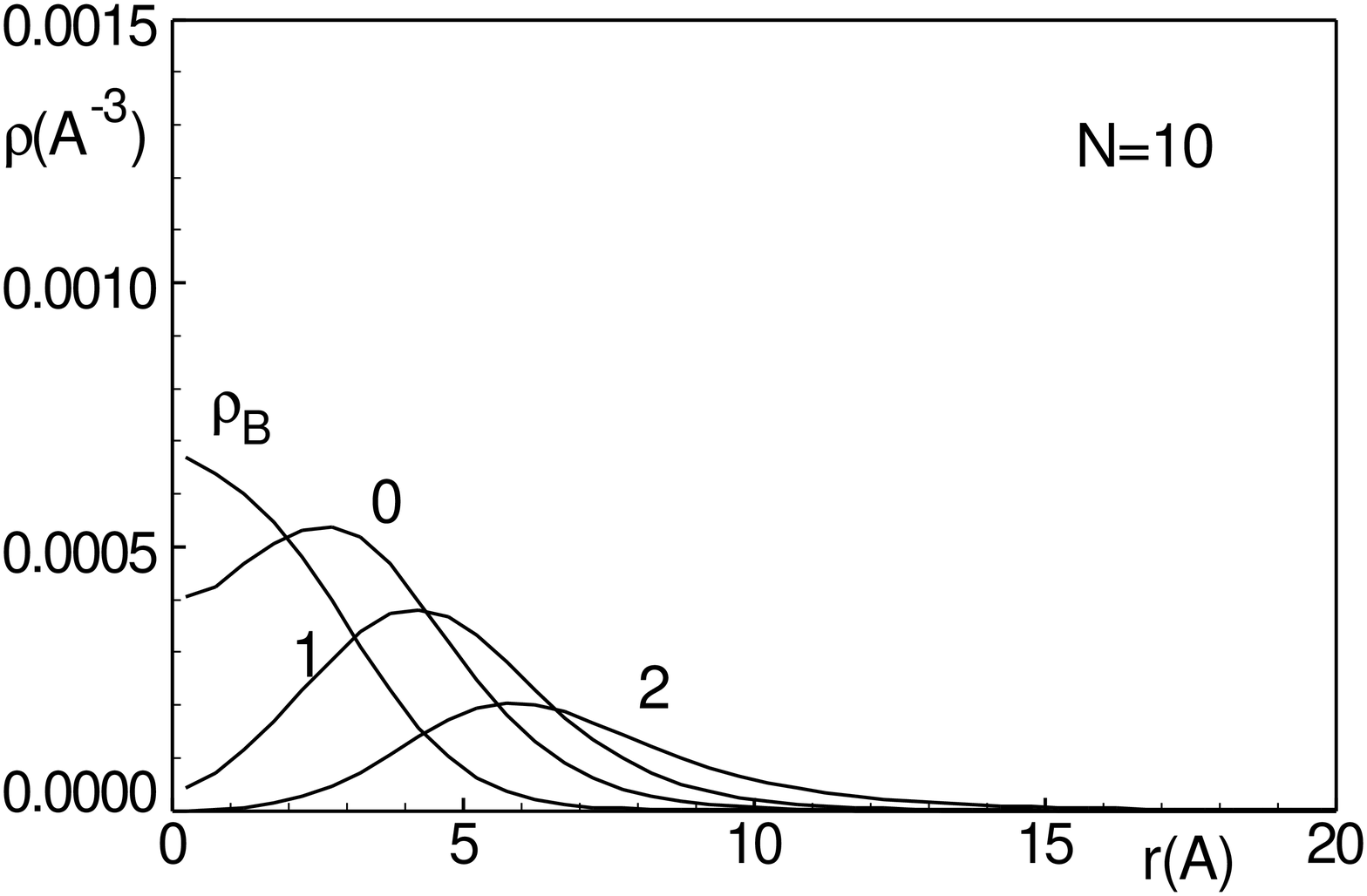}
\hfill
\includegraphics[width=4.25cm]{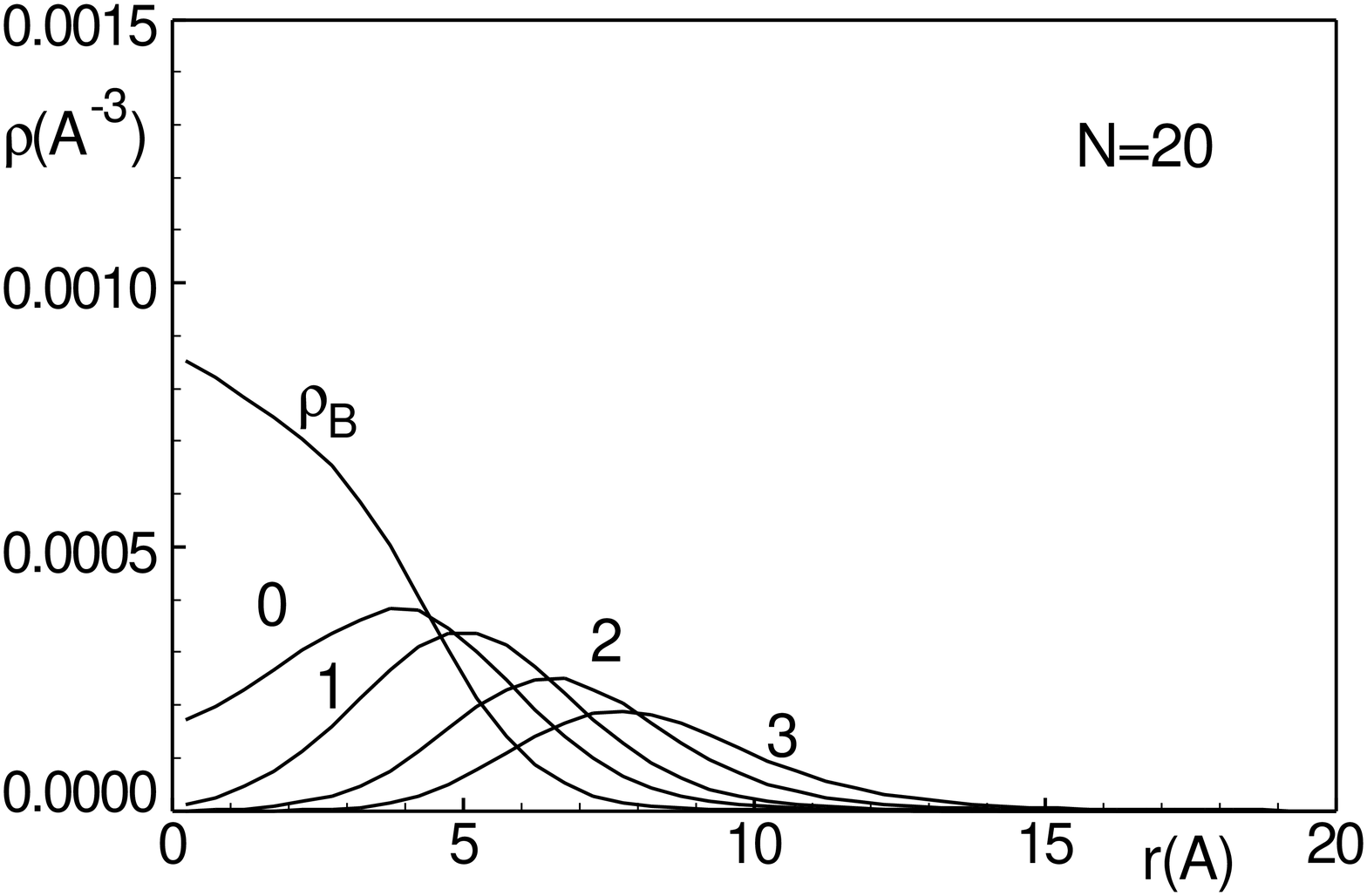}
\includegraphics[width=4.25cm]{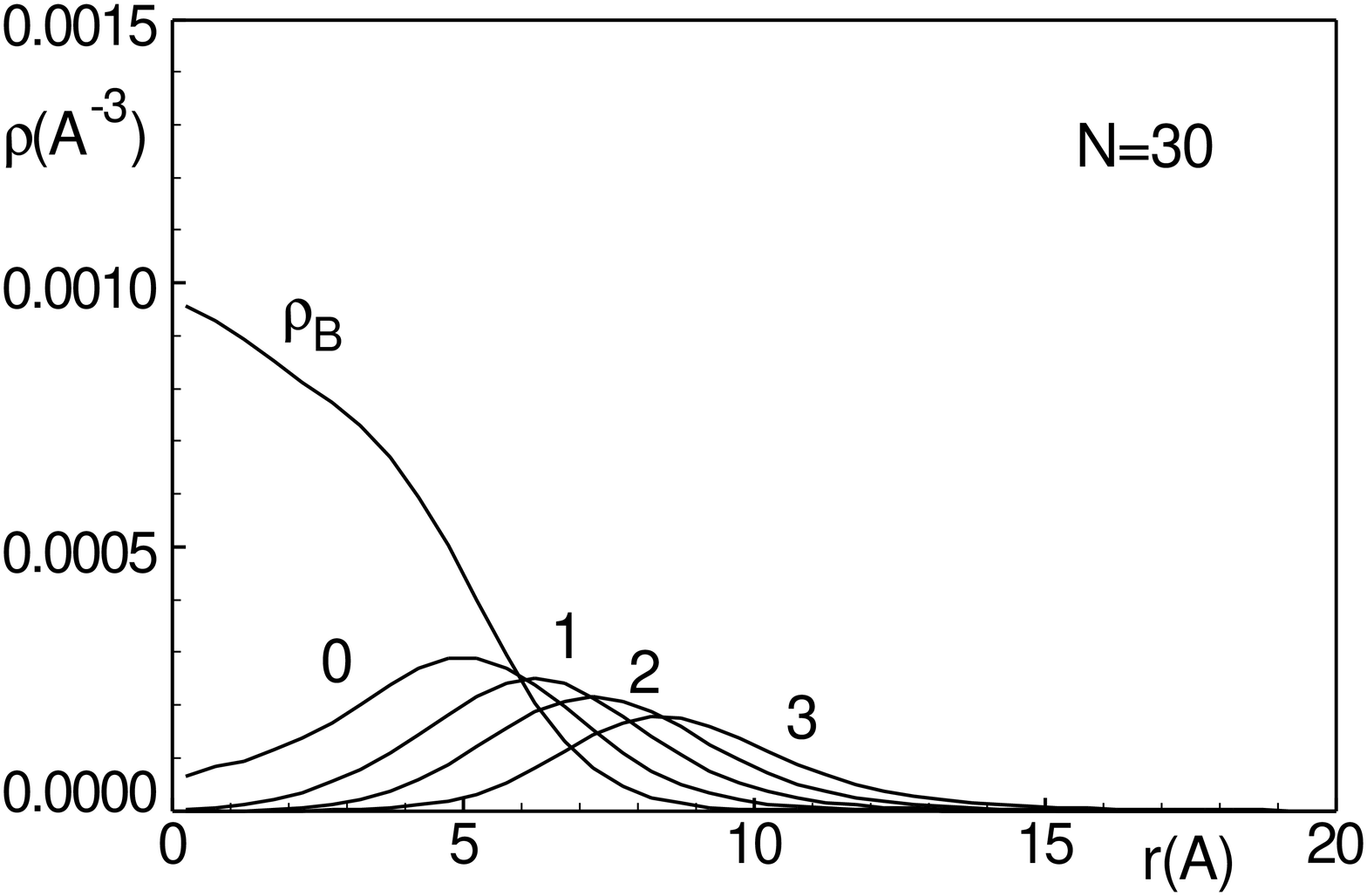}
\hfill
\includegraphics[width=4.25cm]{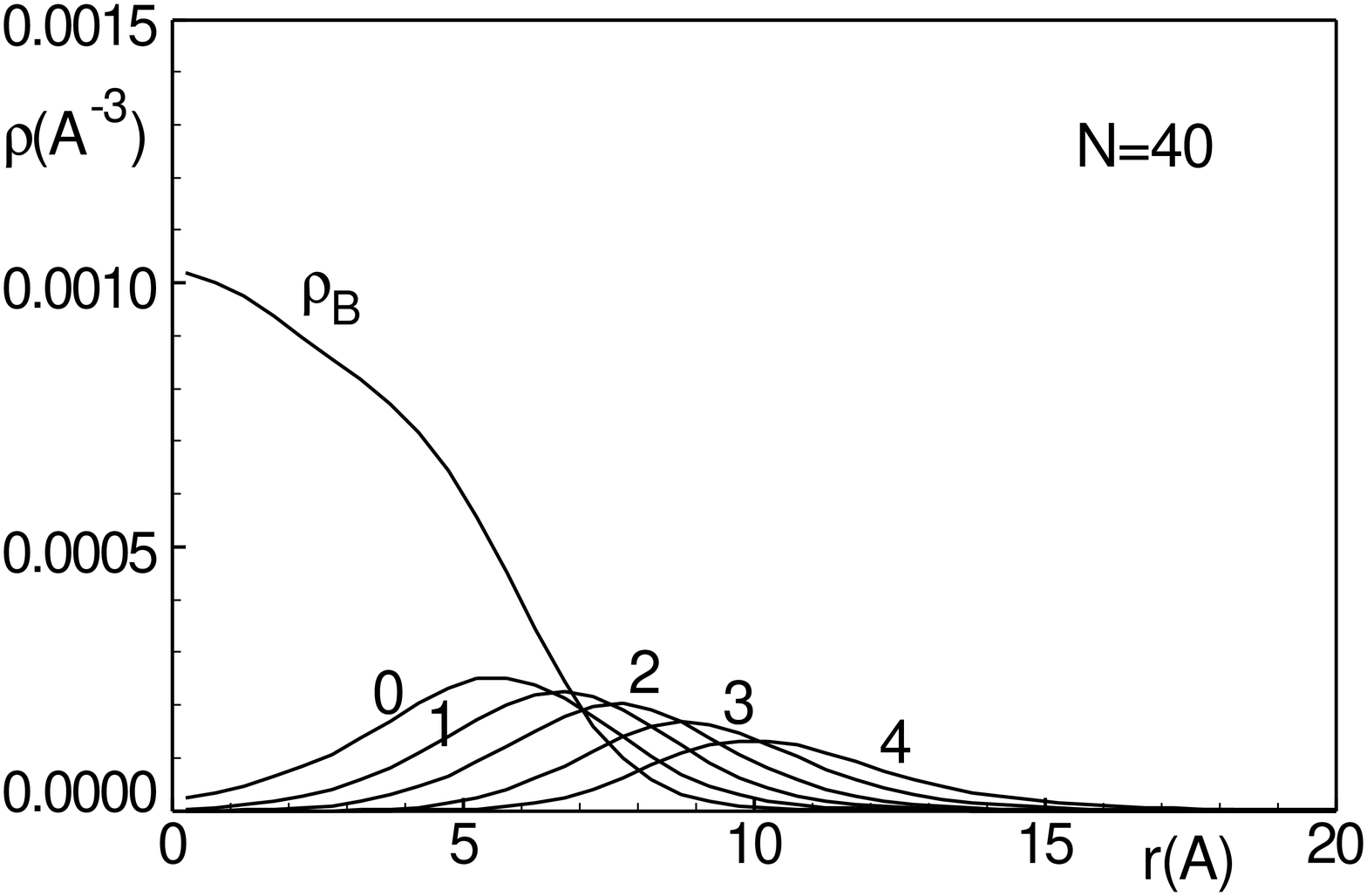}
\end{figure}

A complementary information about the nature of the excitations is provided
by the density distributions of the fermion with respect to the center-of-mass
of the system. Given that we are dealing with non-zero angular momentum states,
to simplify the presentations we show them in  Figure \ref{distributions},
for the bound states of systems with $N = 10$, 20, 30, and 40. No plot of 
radially excited states is given, because the calculation of excitation energies
using sum rules does not allow to obtain the density distributions. In all the 
figures $\rho_B$ corresponds to the boson distribution of the ground state with 
$L=0$, as there are no appreciable differences between the boson distributions 
for drops in other $L$-excited state. This fact supports the model of an 
$^3$He atom in a potential well created by the $^4$He$_N$ atoms, like in the
Lekner approximation.

Table~\ref{sizes} lists the values of the root mean square radii of bosons and
fermions with respect to the center of mass. Only the ground state boson radius
has been displayed in this table  since it is almost independent of the angular 
momentum $L$. The boson radius grows monotonically with the number of bosons, 
following a rough $N^{1/3}$ law, as could be expected. The fermion radii are 
given for the states with angular momentum from $L=0-4$. At fixed $L$, the 
fermion radii increases smoothly with $N$, except for the lowest value 
signaling the threshold of stability. In that case, the fermion radius may be 
abnormally large, indicating that the system is only slightly bound. A flashy 
case is $N=30, L=4$ due to its very large fermion radius. This level is almost 
surely unbound and the DMC algorithm  ejects the fermion far away from the 
center-of-mass of the drop; eventually the $^3$He will move to infinity if the 
random walk were long enough.

\begin{table}[h!]
\caption{Average boson (second column) and fermion (third to seventh columns)
distances (in \AA) to the center-of-mass of the cluster. Empty entries
correspond to unbound systems. }
\label{sizes}
\begin{tabular}{rrrrrrrr}
\hline
$N$ & $r_B$ & $L=0$ & $L=1$ & $L=2$ & $L=3$ & $L=4$ \\
\hline
  5  & 4.67 &  6.16  &    8.04 &      &   &   \\
 10  & 5.16 &  6.79  &    7.87 &    10.45  &    &    \\
 15  & 5.49 &  7.01  &    8.02 &     8.93  &   10.74  &   \\
 20  & 5.86 &  7.50  &    8.12 &     9.19  &   10.25  &    \\
 25  & 6.15 &  7.80  &    8.42 &     9.26  &   10.07  &    \\
 30  & 6.41 &  8.32  &    9.09 &     9.67  &   10.67  &   17.88 \\
 35  & 6.64 &  8.50  &    9.14 &     9.87  &   10.41  &   11.62 \\
 40  & 6.85 &  8.79  &    9.37 &     9.90  &   10.85  &   11.79 \\
 45  & 7.05 &  9.02  &    9.68 &    10.18  &   10.79  &   11.78 \\
 50  & 7.24 &  9.17  &    9.81 &    10.47  &   11.08  &   12.00 \\
\hline
\end{tabular}
\end{table}

The picture which emerges from the densities plot in Fig.~\ref{distributions}
and from the values of radii in Table~\ref{sizes} is the expected one. The
fermion is always located at the surface of the boson cluster, and increasing
the value of the angular momentum produces  the fermion to go away from the
boson cluster.

\begin{figure}[h!]
\caption{Comparison of the boson distributions in \AA$^{-3}$, normalized to 
the number of particles, for the pure bosonic systems with $N$ and $N+1$ and 
the mixed drop with $N$ bosons and one fermion. Two cases are shown, $N=10$ 
and $N=40$. Labels $10$ and $11$ refer to pure bosonic systems, and $10+1$ to 
the doped drop. In the case of $N=40$ the curves are almost indistinguishable.}
\label{test_lekner}
\includegraphics[width=6cm]{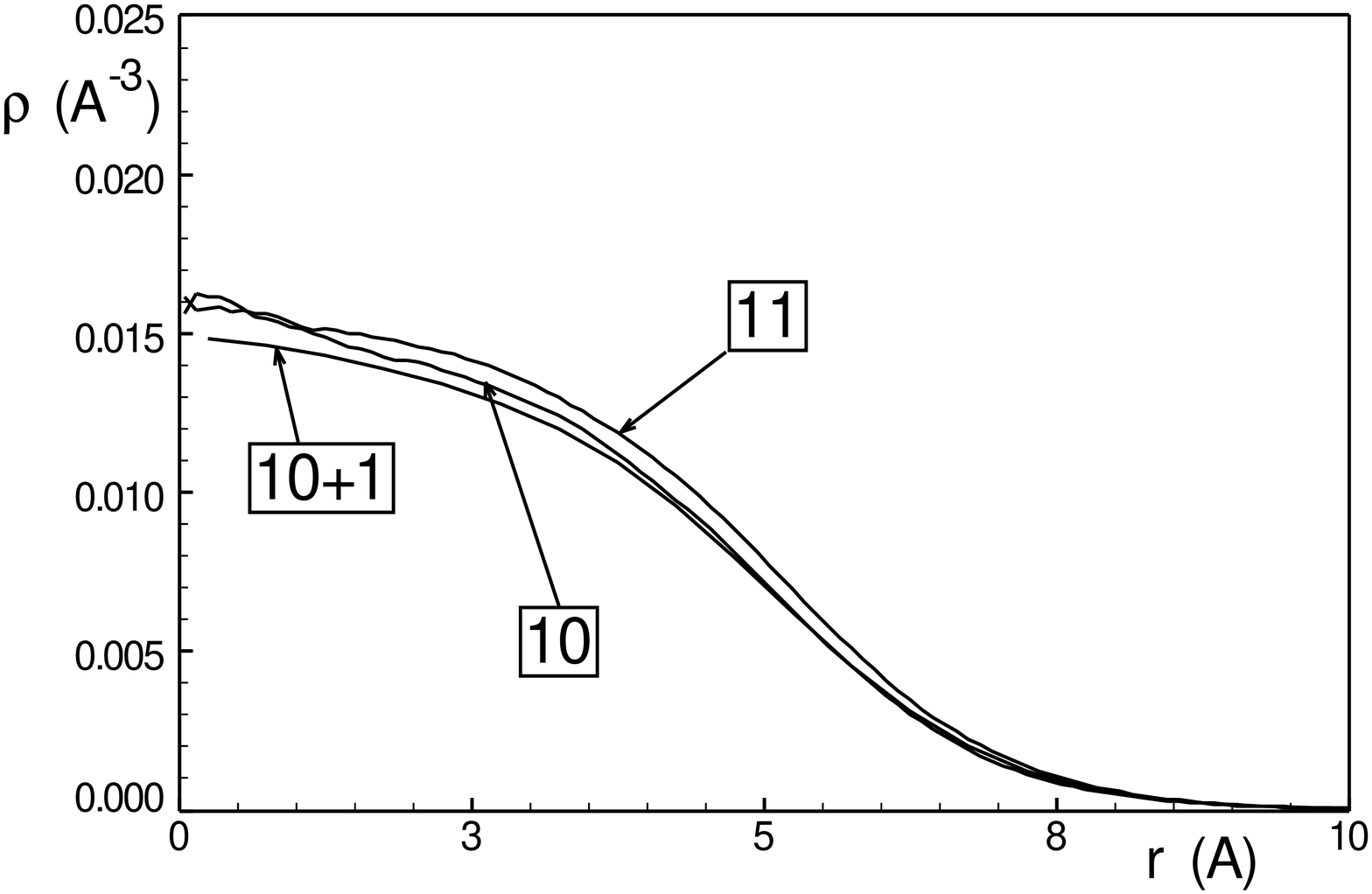}
\includegraphics[width=6cm]{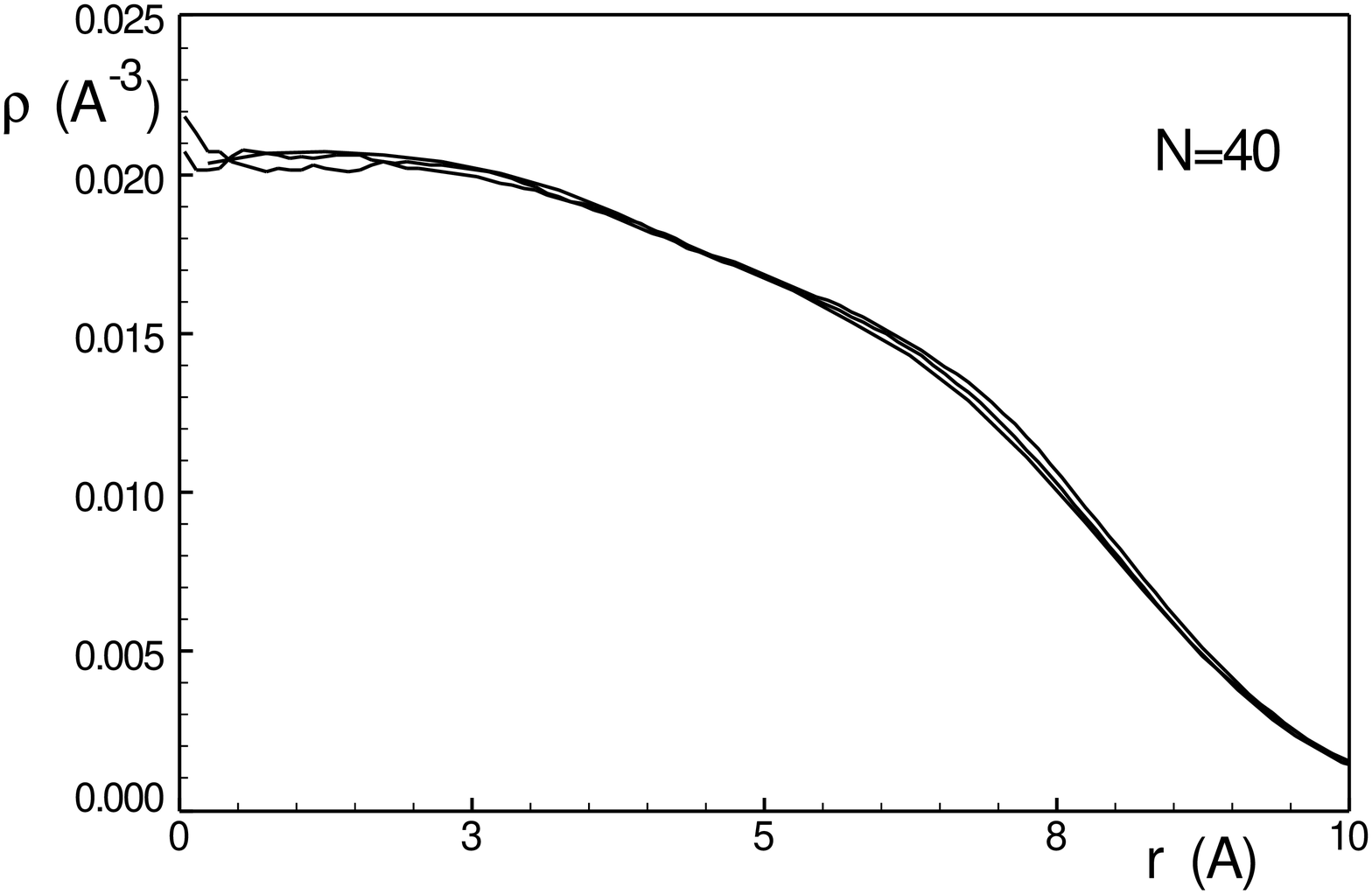}
\end{figure}

To ascertain the goodness of Lekner approximation we have plotted in 
Fig.~\ref{test_lekner} the boson distributions corresponding to two pure $^4$He 
system with $N$ and $N+1$ together with the boson distribution for $N$ bosons 
plus one fermion. The distributions corresponding to $N=40$ are almost 
superimposed, thus revealing the rigidity of the bosonic core in front of the 
addition of a fermion. On the other hand, there are sizable differences between 
the three distributions for $N=10$, revealing the effect of the dopant on the 
bosonic cluster. In other words, there is a weak coupling regime (Lekner 
approximation) for large $N$ but a strong coupling regime for light drops.

\section{Conclusions}

We have permormed DMC simulations, in conjunction with the sum rule method
of Ref.~\cite{Guar01}, to compute the excitation spectrum of a $^3$He impurity
in $^4$He clusters. The rotational levels, namely the lowest energy levels 
within each angular momentum $L$ subspace have been computed by including in 
the guiding function a term $\Phi_L({\bf r}_F - {\bf R}_B)$, generating an 
eigenfunction with good angular momentum quantum numbers. On the other hand, 
the radial excitations have been estimated by computing an optimized upper 
bound obtained with the sum rules of order 0 and 1.

Important results have been obtained for the shell ordering of the $^3$He 
orbitals. First of all, the excitation spectrum  contains a limited set of 
bound excited levels, whose number increases with the number $N$ of bosons. 
Indicating the $n$-th radial excitation of the state with angular momentum $L$ 
with notation $(n+1)L$, using the usual spectroscopic letters for the values 
of $L$, we find the excitation energies to follow a rotational spectrum,
thus suggesting a shell ordering of levels $1s\, 1p\, 1d\, 1f\, 1g \dots$. 
Starting at  $N\approx 20$, the $2s$ radial excited state appears as an 
{\em intruder} within the rotational band. Presumably, at a larger value of $N$,
the $2p$ radial excitation will appear, an so on. The obtained level ordering 
is different from both the $1s\,1p\,1d\,2s\,1f\,2p \dots$ of the 
three-dimensional harmonic oscillator and the $1s\,2s\,1p\,3s\,2p \dots$ 
typical of atoms.

This ordering of levels should be taken into account specially when dealing with
mixed drops with a number of bosons much larger than the number of fermions.
Note however that whereas the DMC algorithm is able to improve the quality of 
the model or importance sampling wave function as far as the bosonic 
correlations are concerned, with respect to the fermionic part it will
maintain the structure of the nodal surfaces. There remains however an
important  question, namely the relevance of these results for pure fermionic
systems or for mixtures with a comparable number of bosons and fermions.
The fermions are expected to play a double role: on the one side, fermions are
creating some kind of self-consistent central field, analogously to the bosons,
and on the other side they are subject to the Pauli principle effects.
So, in a first approximation,  one may assume that the level ordering of
such systems is close to that of the system
$^3{\rm He}\,^4{\rm He}_{N_4+N_3-1}$. However, one should not discard the 
analysis of other alternatives. On the basis of our results it becomes clear 
that previous microscopic calculations of pure fermionic drops as well as of 
mixed drops should be reconsidered, by using an improved importance sampling 
wave function based on realistic shell orderings.

\acknowledgments
This work has been supported by MCyT/FEDER (Spain), grant number BMF2001-0262,
GV (Spain), grant number GV01-216 and MIUR (Italy), cofin-2001025498. One of 
us (RG) acknowledges financial support of the Secretar\'{\i}a de Estado de 
Educaci\'on y Universidades (Spain), Ref. PR2003-0374, as well as DEMOCRITOS 
by his hospitality.

\appendix
\section{The sum rules method applied to excitations of angular momentum $L$}
We have described above the use of the moment method to compute the radial
excitation energies. The method may be also used to determine upper bounds to
the excitation energies of states with angular momentum $L$, as we shall show in
this Appendix. The resulting information will complement the one obtained
directly by the DMC method, and is particularly relevant for the cases $L=1, 2$
where the direct calculation of the excitation energies is affected by a rather
large relative error, being the difference of two large quantities,
specially for large values of $N$.

To obtain upper bounds to the excitation energy of a state of angular momentum
$L\neq 0$ it is convenient to use for the operator $Q$ a form which behaves as
an angular momentum tensor of rank $L$. A simple way is to consider the value
\begin{equation}
\label{exci_L}
Q^{(L)}(R) = f(|{\bf r}_F - {\bf R}_B|) (x_F+iy_F - X_B-i Y_B)^L \, .
\end{equation}
For the function $f$ we have considered a power expansion
\begin{equation}
f(r) = \sum_{n=0} C_n r^n \, ,
\end{equation}
with parameters $C_n$ to be determined after optimization of the upper bound.
To fulfill the requirements which lead to Eq.~(\ref{upper_bound}) we should 
consider the Hermitian part of this operator. In fact, as we are interested in 
excitations of angular momentum $L$, irrespective of the value of the projection
of angular momentum along some fixed axis, a linear combination like
$$
\tilde Q ^{(L)} =  \left( Q^{(L)} + {Q^{(L)}}^{\dagger} \right) / \sqrt{2}
$$
will be adequate. This combination has the advantadge that sum rules $M_0$ and 
$M_1$ are expressed by Eqs.~(\ref{eme0}) and (\ref{simple_m1}), replacing 
operator $Q$ in these expressions with either $Q^{(L)}$ or its Hermitian 
conjugate ${Q^{(L)}}^{\dagger}$.

As in the case of radial excitations, we end up with a generalized eigenvalue 
problem, the required matrix elements of the moment operators being given by
\begin{equation}
{\cal M}^1_{mn} =\frac{\hbar^2}{2\mu}
\langle 0 | \nabla [r^m(x-iy)^L] \cdot \nabla[r^n(x+iy)^L] | 0 \rangle
\end{equation}
and
\begin{equation}
{\cal M}^0_{mn} =
\langle 0 |  r^{m+n}(x^2+y^2)^L | 0 \rangle \, ,
\end{equation}
with ${\bf r} = {\bf r}_F - {\bf R}_B$. Notice that $<0|Q^{(L)}|0>=0$.

As the expectation values are taken with respect to the $L=0$ ground
state, one may take the angular averages of the operators. After some
algebra there results
\begin{equation}
{\cal M}^0_{mn} = \langle 0 | r^{2L+m+n} | 0 \rangle {\cal I}_L
\end{equation}
and
\begin{equation}
{\cal M}^1_{mn}=\frac{\hbar^2}{2\mu}
\langle 0 | r^{2L+m+n-2} | 0 \rangle
[ 2 L^2 {\cal I}_{L-1} + (Lm+Ln+mn) {\cal I}_L ]
\end{equation}
with
\begin{equation}
{\cal I}_L = \frac{1}{2} \int_0^\pi \sin^{2L+1} \theta d\theta
\equiv \frac {L! 2^L}{(2L+1)!!} \, .
\end{equation}

Particularly simple are the expressions for the upper bound when only the
$m=0, n=0$ matrix elements are retained, namely $f(r)$ is taken as a constant,
\begin{equation}
\label{rot_band}
\delta E_L = \frac{\hbar^2}{2 \mu}
\frac{2 \langle r^{2L-2}\rangle}{\langle r^{2L}\rangle}
L(L+\frac{1}{2}) \, ,
\end{equation}
which recalls the naive the rotational model. It is worth mentioning the 
difference between this bound for angular excitations, Eq.~(\ref{rot_band}) 
and the bound obtained for radial excitations, Eq.~(\ref{rad_excit_0}), which 
is manifested specifically in the denominator.

\begin{table}
\label{uppbound}
\caption{Upper bounds to the excitation energy of the lowest energy state
for L=1 to 4}
\begin{tabular}{rrrrrr}
\hline
$N$ & $E_1$ & $E_2$ & $E_3$ & $E_4$ &$\mu _F$\\
\hline
10 & 0.43 & 1.01 & 1.57 & 2.04& 1.14 \\
15 & 0.41 & 1.03 & 1.75 & 2.45& 1.43 \\
20 & 0.36 & 0.92 & 1.59 & 2.27& 1.68 \\
25 & 0.33 & 0.87 & 1.55 & 2.31& 1.85 \\
30 & 0.28 & 0.77 & 1.41 & 2.17& 1.97 \\
35 & 0.28 & 0.74 & 1.34 & 2.06& 2.05 \\
40 & 0.25 & 0.69 & 1.29 & 2.06& 2.14 \\
45 & 0.24 & 0.66 & 1.24 & 1.97& 2.21 \\
50 & 0.23 & 0.64 & 1.20 & 1.92& 2.26 \\
\hline
\end{tabular}
\end{table}

In Table~\ref{uppbound} are given the values obtained for these upper bounds,
after solving the generalized eigenvalue problem, and are quite close to the DMC
excitation energies displayed in Table~\ref{spectrum}. As in the case of radial 
excitations, the sum rules have been calculated by means of mixed matrix 
elements, so that they are not strictly variational.


\begin{thebibliography}{99}
\bibitem{Scho94} W. Sch\"ollkopf and J.P. Toennies,
Science {\bf 266}, 1345 (1994).
\bibitem{Scho96} W. Sch\"ollkopf and J.P. Toennies,
J. Chem. Phys. {\bf 104}, 1155 (1996).
\bibitem{Korn03} O. Kornilov and J.P. Toennies
(private communication).
\bibitem{Guar02} R. Guardiola and J. Navarro,
Phys. Rev. Lett. {\bf 89}, 193401 (2002).
\bibitem{Guar03} R. Guardiola and J. Navarro,
Few-Body Syst. Suppl. {\bf 14}, 223 (2003).
\bibitem{Guar03a} R. Guardiola and J. Navarro,
Phys. Rev. A {\bf 68}, 055201 (2003).
\bibitem{Bres00} D. Bressanini, M. Zavaglia, M. Mella and G. Morosi,
J. Chem. Phys. {\bf 112}, 717 (2000).
\bibitem{Bres02} D. Bressanini, G. Morosi, L. Bertini, M. Mella,
Few-Body Syst. {\bf 31}, 199 (2002).
\bibitem{Bres03} D. Bressanini and G. Morosi,
Phys. Rev. Lett. {\bf 90}, 133401 (2003).
\bibitem{Barr97} M. Barranco, M. Pi, S.M. Gatica, E.S. Hern\'andez and 
J. Navarro,
Phys. Rev. B {\bf 56}, 8997 (1997).
\bibitem{bruch02}
L.W. Bruch, W. Sch\"ollkopf and J.P. Toennies,
J. Chem. Phys. {\bf117}, 1544 (2002).
\bibitem{nava04} J. Navarro, M. Barranco, M. Pi, and A. Poves,
Phys. Rev. A {\bf 69}, in press (2004).
\bibitem{HFD-B}
R.A. Aziz, F.R. McCourt and C.C.K. Wong,
Mol. Phys. {\bf 61}, 1487 (1987).
\bibitem{LM2M2}
R.A. Aziz and M.J. Slaman,
Jour. Chem. Phys. {\bf 94}, 8047 (1991).
\bibitem{TTY}
K.T. Tang, J.P. Toennies and C.L. Yiu,
Phys. Rev. Lett. {\bf 74}, 1546 (1995).
\bibitem{Dalf89} F. Dalfovo,
Z. Phys. D {\bf 14}, 263 (1989).
\bibitem{Beli94} A. Beli\u c, F. Dalfovo, S. Fantoni and S. Stringari,
Phys. Rev. B {\bf 49}, 15253 (1994).
\bibitem{Krot01} E. Krotscheck and R. Zillich,
J. Chem. Phys. {\bf 115}, 10161 (2001).
\bibitem{Lekn70} J. Lekner,
Philos. Mag. {\bf 22}, 669 (1970).
\bibitem{Guar99} R. Guardiola, M. Portesi and J. Navarro,
Phys. Rev. B {\bf 60}, 6288 (1999).
\bibitem{Guar00} R. Guardiola and J. Navarro,
Phys. Rev. Lett {\bf 84}, 1144 (2000).
\bibitem{Guar00a} R. Guardiola,
Phys. Rev. B {\bf 62}, 3416 (2000).
\bibitem{Guar01} R. Guardiola, J. Navarro and M. Portesi,
Phys. Rev. B {\bf 63}, 224519 (2001).
\bibitem{Chin90} S.A. Chin and E. Krotscheck,
Phys. Rev. Lett. {\bf 65}, 2658 (1990).
\bibitem{Chin92} S.A. Chin and E. Krotscheck,
Phys. Rev. B {\bf 45}, 852 (1992).
\bibitem{Kris90} M.V. Rama Krisna and K.B. Waley,
J. Chem. Phys. {\bf 93}, 6738 (1990).
\bibitem{Reyn82} P.J. Reynolds, D.M. Ceperley, B.J. Alder and W.A. Lester Jr.,
J. Chem. Phys. {\bf 77}, 5593 (1982).
\bibitem{Mosk82} J.W. Moskowitz, K.E. Schmidt, M.A. Lee and H.M. Kalos,
J. Chem. Phys. {\bf 77}, 349 (1982).
\bibitem{Vrbi86} J. Vrbik and S.M. Rothstein,
J. Comput. Phys. {\bf 63}, 130 (1986).
\bibitem{Chin90a}  Siu A. Chin,
Phys. Rev. A {\bf 42}, 6991 (1990).
\bibitem{Cepe81} D.M. Ceperley,
in {\em Recent Progress in Many-Body theories}, edited by J.G. Zabolitzky,
M. de Llano, M. Fortes and J.W. Clark, Lect. Not. Phys. {\bf 142}, 262
(Springer Verlag, Berlin 1981)
\end{thebibliography}
\end{document}